%% file: 2020-online-counterfactual-evaluation.tex
\pdfoutput=1

\documentclass[sigconf,natbib]{acmart}

\input{packages}
\input{meta}

\input{definitions}
\input{authors}

\renewcommand{\arraystretch}{0.0}

\title[Taking the Counterfactual Online: Efficient and Unbiased Online Evaluation for Ranking]{Taking the Counterfactual Online:\\ Efficient and Unbiased Online Evaluation for Ranking}

\allowdisplaybreaks

\begin{document}

\begin{abstract}
Counterfactual evaluation can estimate \ac{CTR} differences between ranking systems based on historical interaction data, while mitigating the effect of position bias and item-selection bias.
We introduce the novel \acf{LogOpt}, which optimizes the policy for logging data so that the counterfactual estimate has minimal variance.
As minimizing variance leads to faster convergence, \ac{LogOpt} increases the data-efficiency of counterfactual estimation.
\ac{LogOpt} turns the counterfactual approach -- which is indifferent to the logging policy -- into an online approach, where the algorithm decides what rankings to display.
We prove that, as an online evaluation method, \ac{LogOpt} is unbiased w.r.t.\ position and item-selection bias, unlike existing interleaving methods.
Furthermore, we perform large-scale experiments by simulating comparisons between \emph{thousands} of rankers.
Our results show that while interleaving methods make systematic errors, \ac{LogOpt} is as efficient as interleaving without being biased.
\end{abstract}

\maketitle

\acresetall

\input{sections/01-introduction}
\input{sections/02-ctr-estimators}
\input{sections/03-related-work}
\input{sections/04-method}

\input{sections/05-experimental-setup}
\input{sections/06-results}

\input{sections/07-conclusion}


\subsection*{Acknowledgements}
We want to thank the anonymous reviewers for their feedback.
This research was partially supported by the Netherlands Organisation for Scientific Research (NWO) under project nr 612.001.551 and by the Innovation Center for AI (ICAI).
All content represents the opinion of the authors, which is not necessarily shared or endorsed by their respective employers and/or sponsors.

\subsection*{Reproducibility}
Our experimental implementation is publicly available at \url{https://github.com/HarrieO/2020ictir-evaluation}.


\bibliographystyle{ACM-Reference-Format}
\bibliography{references}

\appendix
\input{sections/08-appendix}

\end{document}

%% file: packages.tex
\PassOptionsToPackage{dvipsnames}{xcolor}
\PassOptionsToPackage{pdflatex}{graphicx}

\usepackage{dsfont}
\usepackage{acronym}
\usepackage{amsmath}
\usepackage[noend]{algorithmic}
\usepackage{algorithm}
\usepackage{bbm}
\usepackage{beramono}
\usepackage{bm}
\usepackage{bmpsize}
\usepackage{booktabs}
\usepackage[skip=0pt]{caption}
\usepackage{cleveref}
\usepackage[T1]{fontenc}
\usepackage{graphicx}
\usepackage{hyperref}
\usepackage{listings}
\usepackage{multirow}
\usepackage{natbib}
\usepackage{subcaption}
\usepackage{tabularx}
\usepackage[dvipsnames]{xcolor}
\usepackage{enumerate}
\usepackage[inline]{enumitem}
\usepackage{numprint}
\usepackage[normalem]{ulem}

%% file: meta.tex
\copyrightyear{2020}
\acmYear{2020}
\setcopyright{acmlicensed}\acmConference[ICTIR '20]{Proceedings of the 2020 ACM SIGIR International Conference on the Theory of Information Retrieval}{September 14--17, 2020}{Virtual Event, Norway}
\acmBooktitle{Proceedings of the 2020 ACM SIGIR International Conference on the Theory of Information Retrieval (ICTIR '20), September 14--17, 2020, Virtual Event, Norway}
\acmPrice{15.00}
\acmDOI{10.1145/3409256.3409820}
\acmISBN{978-1-4503-8067-6/20/09}

%% file: definitions.tex
\acrodef{IR}{Information Retrieval}
\acrodef{LTR}{Learning to Rank}
\acrodef{ARP}{Average Relevance Position}
\acrodef{DCG}{Discounted Cumulative Gain}
\acrodef{EM}{Expectation Maximization}
\acrodef{CTR}{Click-Through-Rate}
\acrodef{IPS}{Inverse-Propensity-Scoring}
\acrodef{LogOpt}{Logging-Policy Optimization Algorithm}

\DeclareMathOperator*{\argmin}{arg\,min}

\allowdisplaybreaks

%% file: authors.tex
\author{Harrie Oosterhuis}
\affiliation{%
	\institution{University of Amsterdam}
	\city{Amsterdam}
	\country{The Netherlands}
}
\email{oosterhuis@uva.nl}
 
\author{Maarten de Rijke}
\orcid{0000-0002-1086-0202}
\affiliation{
 \institution{University of Amsterdam \& Ahold Delhaize}
 \city{Amsterdam}
 \country{The Netherlands}
}
\email{derijke@uva.nl}

%% file: sections/01-introduction.tex

\section{Introduction}
\label{sec:intro}

Evaluation is essential for the development of search and recommendation systems~\citep{hofmann2016online, kohavi2017online}.
Before any ranking model is widely deployed it is important to first verify whether it is a true improvement over the currently-deployed model.
A traditional way of evaluating relative differences between systems is through A/B testing, where part of the user population is exposed to the current system (``control") and the rest to the altered system (``treatment") during the same time period.
Differences in behavior between these groups can then indicate if the alterations brought improvements, e.g.\ if the treatment group showed a higher \ac{CTR} or more revenue was made with this system~\citep{chapelle2012large}.

Interleaving has been introduced in \ac{IR} as a more efficient alternative to A/B testing~\citep{joachims2003evaluating}.
Interleaving algorithms take the rankings produced by two ranking systems, and for each query create an interleaved ranking by combining the rankings from both systems.
Clicks on the interleaved rankings directly indicate relative differences.
Repeating this process over a large number of queries and averaging the results, leads to an estimate of which ranker would receive the highest \ac{CTR}~\citep{hofmann2013fidelity}.
Previous studies have found that interleaving requires fewer interactions than A/B testing, which enables them to make consistent comparisons in a much shorter timespan~\citep{schuth2015predicting, chapelle2012large}.

More recently, counterfactual evaluation for rankings has been proposed by \citet{joachims2017unbiased} to evaluate a ranking model based on clicks gathered using a different model.
By correcting for the position bias introduced during logging, the counterfactual approach can unbiasedly estimate the \ac{CTR} of a new model on historical data.
To achieve this, counterfactual evaluation makes use of \ac{IPS}, where clicks are weighted inversely to the probability that a user examined them during logging~\citep{wang2016learning}.
A big advantage compared to interleaving and A/B testing, is that counterfactual evaluation does not require online interventions.

In this paper, we show that no existing interleaving method is truly unbiased: they are not guaranteed to correctly predict which ranker has the highest \ac{CTR}.
On two different industry datasets, we simulate a total of 1,000 comparisons between 2,000 different rankers.
In our setup, interleaving methods converge on the wrong answer for at least 2.2\% of the comparisons on both datasets.
A further analysis shows that existing interleaving methods are unable to reliably estimate \ac{CTR} differences around 1\% or lower.
Therefore, in practice these systematic errors are expected to impact situations where rankers with a very similar \ac{CTR} are compared.

We propose a novel online evaluation algorithm: \acf{LogOpt}.
\ac{LogOpt} extends the existing unbiased counterfactual approach, and turns it into an online approach.
\ac{LogOpt} estimates which rankings should be shown to the user, so that the variance of its \ac{CTR} estimate is minimized.
In other words, it attempts to learn the logging-policy that leads to the fastest possible convergence of the counterfactual estimation.
Our experimental results indicate that our novel approach is as efficient as any interleaving method or A/B testing, without having a systematic error.
As predicted by the theory, we see that the estimates of our approach converge on the true \ac{CTR} difference between rankers.
Therefore, we have introduced the first online evaluation method that combines high efficiency with unbiased estimation.

The main contributions of this work are:
\begin{enumerate}[leftmargin=*,nosep]
\item The first logging-policy optimization method for minimizing the variance in counterfactual \ac{CTR} estimation.
\item The first unbiased online evaluation method that is as efficient  as state-of-the-art interleaving methods.
\item A large-scale analysis of existing online evaluation methods that reveals a previously unreported bias in interleaving methods.
\end{enumerate}

%% file: sections/02-ctr-estimators.tex

\section{Preliminaries: Ranker Comparisons}

The overarching goal of ranker evaluation is to find the ranking model that provides the best rankings.
For the purposes of this paper, we will define the quality of a ranker in terms of the number of clicks it is expected to receive.
Let $R$ indicate a ranking and let $\mathbbm{E}[\text{CTR}(R)] \in \mathbb{R}_{\geq0}$ be the expected number of clicks a ranking receives after being displayed to a user.
We consider ranking $R_1$ to be better than $R_2$ if in expectation it receives more clicks: $\mathbbm{E}[\text{CTR}(R_1)] > \mathbbm{E}[\text{CTR}(R_2)]$.
We will represent a ranking model by a policy $\pi$, with $\pi(R \mid q)$ as the probability that $\pi$ displays $R$ for a query $q$.
With $P(q)$ as the probability of a query $q$ being issued, the expected number of clicks received under a ranking model $\pi$ is:
\begin{equation}
\mathbbm{E}[\text{CTR}(\pi)]
=
\sum_{q} P(q) \sum_{R} \mathbbm{E}[\text{CTR}(R)] \pi(R \mid q).
\end{equation}
Our goal is to discover the $\mathbbm{E}[\text{CTR}]$ difference between two policies:
\begin{equation}
\Delta(\pi_1, \pi_2) 
= 
\mathbbm{E}[\text{CTR}(\pi_1)] - \mathbbm{E}[\text{CTR}(\pi_2)].
\label{eq:ctrdifference}
\end{equation}
We recognize that to correctly identify if one policy is better than another, we merely need a corresponding binary indicator:
\begin{equation}
\Delta_{bin}(\pi_1, \pi_2) 
= 
\text{sign}\big(
\Delta(\pi_1, \pi_2) 
\big).
\label{eq:binctrdifference}
\end{equation}
However, in practice the magnitude of the differences can be very important, for instance, if one policy is much more computationally expensive while only having a slightly higher $\mathbbm{E}[\text{CTR}]$, it may be preferable to use the other in production.
Therefore, estimating the absolute $\mathbbm{E}[\text{CTR}]$ difference is more desirable in practice.


\subsection{User Behavior Assumptions}
\label{sec:preliminary:assumptions}

Any proof regarding estimators using user interactions must rely on assumptions about user behavior.
In this paper, we assume that only two forms of interaction bias are at play: position bias and item-selection bias.

Users generally do not examine all items that are displayed in a ranking but only click on examined items~\citep{chuklin2015click}.
As a result, a lower probability of examination for an item also makes it less likely to be clicked.
Position bias assumes that only the rank determines the probability of examination~\citep{craswell2008experimental}.
Furthermore, we will assume that given an examination only the relevance of an item determines the click probability.
Let $c(d) \in \{0,1\}$ indicate a click on item $d$ and $o(d) \in \{0,1\}$ examination by the user.
Then these assumptions result in the following assumed click probability:
\begin{equation}
\begin{split}
P(c(d) = 1 \mid R, q)
&= 
P(o(d) = 1 \mid R)
P(c(d) = 1 \mid o(d) = 1, q)
\\
&= 
\theta_{\text{rank}(d \mid R)}
\zeta_{d, q}.
\end{split}
\end{equation}
Here $\text{rank}(d \mid R)$ indicates the rank of $d$ in $R$; for brevity we use $\theta_{\text{rank}(d \mid R)}$ to denote the examination probability -- $\theta_{\text{rank}(d \mid R)} = P(o(d) = 1 \mid R)$ -- and $\zeta_{d, q}$ for the conditional click probability -- $\zeta_{d, q} = P(c(d) = 1 \mid o(d) = 1, q)$.


We also assume that item-selection bias is present; this type of bias is an extreme form of position bias that results in zero examination probabilities for some items~\citep{ovaisi2020correcting, oosterhuis2020topkrankings}.
This bias is unavoidable in top-$k$ ranking settings, where only the $k \in \mathbb{N}_{> 0}$ highest ranked items are displayed.
Consequently, any item beyond rank $k$ cannot be observed or examined by the user:
$
\forall r \in \mathbb{N}_{> 0}\, (r > k \rightarrow \theta_{r} = 0)
$.
The distinction between item-selection bias and position bias is important because the original counterfactual evaluation method~\citep{joachims2017unbiased} is only able to correct for position bias when no item-selection bias is present~\citep{ovaisi2020correcting, oosterhuis2020topkrankings}. 

Based on these assumptions, we can now formulate the expected \ac{CTR} of a ranking:
\begin{equation}
\mathbbm{E}[\text{CTR}(R)] = \sum_{d \in R} P(c(d) = 1 \mid R, q) = \sum_{d \in R} \theta_{\text{rank}(d \mid R)}
\zeta_{d, q}.
\end{equation}
While we assume this model of user behavior, its parameters are still assumed unknown.
Therefore, the methods in this paper will have to estimate $\mathbbm{E}[\text{CTR}]$ without prior knowledge of $\theta$ or $\zeta$.

\subsection{Goal: \ac{CTR}-Estimator Properties}
\label{sec:preliminary:properties}

Recall that our goal is to estimate the \ac{CTR} difference between rankers~(Eq.~\ref{eq:ctrdifference}); online evaluation methods do this based on user interactions.
Let $\mathcal{I}$ be the set of available user interactions, it contains $N$ tuples of a single (issued) query $q_i$, the corresponding displayed ranking $R_i$, and the observed user clicks $c_i$:
$\mathcal{I} = \{(q_i, R_i, c_i)\}^N_{i=1}$.
Each evaluation method has a different effect on what rankings will be displayed to users.
Furthermore, each evaluation method converts each interaction into a single estimate using some function $f$:
$x_i = f(q_i, R_i, c_i)$, the final estimate is simply the mean over these estimates:
$\hat{\Delta}(\mathcal{I}) = \frac{1}{N} \sum_{i=1}^N x_i = \frac{1}{N} \sum_{i=1}^N f(q_i, R_i, c_i)$.
This description fits all existing online and counterfactual evaluation methods for rankings.
Every evaluation method uses a different function $f$ to convert interactions into estimates; moreover, online evaluation methods also decide which rankings $R$ to display when collecting $\mathcal{I}$. 
These two choices result in different estimators.
Before we discuss the individual methods, we briefly introduce the three properties we desire of each estimator: consistency, unbiasedness and variance.

\textbf{Consistency} -- an estimator is \emph{consistent} if it converges as $N$ increases.
All existing evaluation methods are consistent as their final estimates are means of bounded values. 

\textbf{Unbiasedness} -- an estimator is \emph{unbiased} if its estimate is equal to the true \ac{CTR} difference in expectation:
\begin{equation}
\text{Unbiased}(\hat{\Delta}) \Leftrightarrow  \mathbbm{E}\big[\hat{\Delta}(\mathcal{I})\big] = \Delta(\pi_1, \pi_2).
\end{equation}
If an estimator is both consistent and unbiased it is guaranteed to converge on the true $\mathbbm{E}[\text{CTR}]$ difference.

\textbf{Variance} -- the \emph{variance} of an estimator is the expected squared deviation between a single estimate $x$ and the mean $\hat{\Delta}(X)$:
\begin{equation}
\text{Var}(\hat{\Delta}) =  \mathbbm{E}\Big[ \big(x -  \mathbbm{E}[\hat{\Delta}(\mathcal{I})]\big)^2  \Big].
\end{equation}
Variance affects the rate of convergence of an estimator; for fast convergence it should be as low as possible.

In summary, our goal is to find an estimator, for the \ac{CTR} difference between two ranking models, that is consistent, unbiased and has minimal variance.

%% file: sections/03-related-work.tex

\section{Existing Online and Counter- factual Evaluation Methods}
We describe three families of online and counterfactual evaluation methods for ranking.

\subsection{A/B Testing}
A/B testing is a well established form of online evaluation to compare a system A with a system B~\citep{kohavi2017online}.
Users are randomly split into  two groups and during the same time period each group is exposed to only one of the systems.
In expectation, the only factor that differs between the groups is the exposure to the different systems.
Therefore, by comparing the behavior of each user group, the relative effect each system has can be evaluated.

We will briefly show that A/B testing is unbiased for $\mathbbm{E}[\text{CTR}]$ difference estimation.
For each interaction either $\pi_1$ or $\pi_2$ determines the ranking, let $A_i \in \{1, 2\}$ indicate the assignment and $A_i \sim P(A)$.
Thus, if $A_i = 1$, then $R_i \sim \pi_1(R \mid q)$ and if $A_i = 2$, then $R_i \sim \pi_2(R \mid q)$.
Each interaction $i$ is converted into a single estimate $x_i$ by $f_{\text{A/B}}$:
\begin{equation}
x_i =  f_{\text{A/B}}(q_i, R_i, c_i) = \left(\frac{\mathbbm{1}[A_i = 1]}{P(A = 1)} - \frac{\mathbbm{1}[A_i = 2]}{P(A = 2)}\right) \sum_{d \in R_i} c_i(d).
\end{equation}
Abbreviating $f_{\text{A/B}}(q_i, R_i, c_i)$ as $f_{\text{A/B}}(\ldots)$, we can prove that A/B testing is unbiased, since in expectation each individual estimate is equal to the \ac{CTR} difference:
\begin{align}
\mathbbm{E}[f_{\text{A/B}}(\ldots)] &=
\sum_q P(q) \bigg(\frac{ P(A = 1) \sum_R \pi_1(R \mid q) \mathbbm{E}[\text{CTR}(R)] }{P(A = 1)} 
\nonumber \\
& \qquad \qquad \quad - \frac{ P(A = 2) \sum_R \pi_2(R \mid q) \mathbbm{E}[\text{CTR}(R)] }{P(A = 2)} 
\bigg)
\nonumber \\
&= \sum_q P(q) \sum_R \mathbbm{E}[\text{CTR}(R)]\big(\pi_1(R \mid q) - \pi_2(R \mid q)\big)
\nonumber \\
&= \mathbbm{E}[\text{CTR}(\pi_1)] - \mathbbm{E}[\text{CTR}(\pi_2)] = \Delta(\pi_1, \pi_2).
\end{align}
Variance is harder to evaluate without knowledge of $\pi_1$ and $\pi_2$.
Unless $\Delta(\pi_1, \pi_2) = 0$, some variance is unavoidable since A/B testing alternates between estimating $\ac{CTR}(\pi_1)$ and $\ac{CTR}(\pi_2)$.

\subsection{Interleaving}
\label{sec:related:interleaving}
Interleaving methods were introduced specifically for evaluation in ranking, as a more efficient alternative to A/B testing~\citep{joachims2003evaluating}.
After a query is issued, they take the rankings of two competing ranking systems and combine them into a single interleaved ranking.
Any clicks on the interleaved ranking can be interpreted as a preference signal between either ranking system.
Thus, unlike A/B testing, interleaving does not estimate the \ac{CTR} of individual systems but a relative preference; the idea is that this allows it to be more efficient than A/B testing.

Each interleaving method attempts to use randomization to counter position bias, without deviating too much from the original rankings so as to maintain the user experience~\citep{joachims2003evaluating}.
\emph{Team-draft interleaving} (TDI) randomly selects one ranker to place their top document first, then the other ranker places their top (unplaced) document next~\citep{radlinski2008does}.
Then it randomly decides the next two documents, and this process is repeated until all documents are placed in the interleaved ranking.
Clicks on the documents are attributed to the ranker that placed them.
The ranker with the most attributed clicks is inferred to be preferred by the user.
\emph{Probabilistic interleaving} (PI) treats each ranking as a probability distribution over documents; at each rank a distribution is randomly selected and a document is drawn from it~\citep{hofmann2011probabilistic}.
After clicks have been received, probabilistic interleaving computes the expected number of clicks documents per ranking system to infer preferences.
\emph{Optimized interleaving} (OI) casts the randomization as an optimization problem, and displays rankings so that if all documents are equally relevant no preferences are found~\citep{radlinski2013optimized}.

While every interleaving method attempts to deal with position bias, none is unbiased according to our definition (Section~\ref{sec:preliminary:properties}).
This may be confusing because previous work on interleaving makes claims of unbiasedness~\citep{hofmann2013fidelity, radlinski2013optimized, hofmann2011probabilistic}. 
However, they use different definitions of the term.
More precisely, TDI, PI, and OI provably converge on the correct outcome if all documents are equally relevant~\citep{hofmann2013fidelity, radlinski2013optimized, hofmann2011probabilistic, radlinski2008does}.
Moreover, if one assumes binary relevance and $\pi_1$ ranks all relevant documents equal to or higher than $\pi_2$, the binary outcome of PI and OI is proven to be correct in expectation~\citep{radlinski2013optimized, hofmann2013fidelity}.
However, beyond the confines of these unambiguous cases, we can prove that these methods do not meet our definition of unbiasedness: for every method one can construct an example where it converges on the incorrect outcome.
The rankers $\pi_1$, $\pi_2$ and position bias parameters $\theta$ can be chosen so that in expectation the wrong (binary) outcome is estimated; see Appendix~A for a proof for each of the three interleaving methods. 
Thus, while more efficient than A/B testing, interleaving methods make systematic errors in certain circumstances and thus should not be considered to be unbiased w.r.t.\ \ac{CTR} differences.

We note that the magnitude of the bias should also be considered.
If the systematic error of an interleaving method is minuscule while the efficiency gains are very high, it may still be very useful in practice.
Our experimental results (Section~\ref{sec:biasinterleaving}) reveal that the systematic error of all interleaving methods becomes very high when comparing systems with a \ac{CTR} difference of 1\% or smaller.

\subsection{Counterfactual Evaluation}
\label{sec:related:counterfactual}

Counterfactual evaluation is based on the idea that if certain biases can be estimated well, they can also be adjusted
~\citep{joachims2017accurately, wang2016learning}.
While estimating relevance is considered the core difficulty of ranking evaluation, estimating the position bias terms $\theta$ is very doable.
By randomizing rankings, e.g., by swapping pairs of documents~\citep{joachims2017accurately} or exploiting data logged during A/B testing~\citep{agarwal2019estimating}, differences in \ac{CTR} for the same item on different positions can be observed directly.
Alternatively, using \ac{EM} optimization~\citep{wang2018position} or a dual learning objective~\citep{ai2018unbiased}, position bias can be estimated from logged data as well.
Once the bias terms $\theta$ have been estimated, logged clicks can be weighted so as to correct for the position bias during logging.
Hence, counterfactual evaluation can work with historically logged data.
Existing counterfactual evaluation algorithms do not dictate which rankings should be displayed during logging: they do not perform interventions and thus we do not consider them to be online methods.

Counterfactual evaluation assumes that the position bias $\theta$ and the logging policy $\pi_0$ are known, in order to correct for both position bias and item-selection bias.
Clicks are gathered with $\pi_0$ which decides which rankings are displayed to the user.
We follow \citet{oosterhuis2020topkrankings} and use as propensity scores the probability of observance in expectation over the displayed rankings:
\begin{equation}
\begin{split}
\rho(d \mid q) & = \mathbbm{E}_R\big[P(o(d) = 1 \mid R) \mid \pi_0\big] \\
& \textstyle= \sum_R \pi_0(R \mid q) P(o(d) = 1 \mid  R).
\label{eq:prop}
\end{split}
\end{equation}
Then we use $\lambda(d \mid \pi_1, \pi_2)$ to indicate the difference in observance probability under $\pi_1$ or $\pi_2$:
\begin{align}
\lambda(d \,|\, \pi_1, \pi_2)
&= \mathbbm{E}_R\big[P(o(d) = 1 \,|\, R) \,|\, \pi_1\big] - \mathbbm{E}_R\big[P(o(d) = 1 \,|\, R) \,|\, \pi_2\big] \nonumber
\\
&= \sum_R \theta_{\text{rank}(d \mid R)}\big(\pi_1(R \mid q_i)- \pi_2(R \mid q_i)\big).
\label{eq:lambda}
\end{align}
Then, the \ac{IPS} estimate function is formulated as:
\begin{equation}
x_i =  f_{\text{IPS}}(q_i, R_i, c_i) = \sum_{d :  \rho(d | q_i) > 0} \frac{c_i(d) }{\rho(d \,|\, q_i)}\lambda(d \mid \pi_1, \pi_2).
\label{eq:ipsestimator}
\end{equation}
Each click is weighted inversely to its examination probability, but items with a zero probability: $\rho(d \,|\, q_i) = 0$ are excluded.
We note that these items can never be clicked: $\forall q, d \, (\rho(d \,|\, q) = 0 \rightarrow c(d) = 0$).
Before we prove unbiasedness, we note that given $\rho(d \,|\, q_i) > 0$:
\begin{equation}
\begin{split}
\mathbbm{E}\bigg[\frac{c(d)}{\rho(d \,|\, q)}\bigg]
&=
\frac{\sum_R \pi_0(R \mid q) \theta_{\text{rank}(d \mid R)}
\zeta_{d, q}}{\rho(d \,|\, q_i)}
\\
&=
\frac{\sum_R \pi_0(R \mid q) \theta_{\text{rank}(d \mid R)}
}{\sum_{R'} \pi_0(R' \,|\,  q) \theta_{\text{rank}(d \mid R')}} \zeta_{d, q} = \zeta_{d, q}.
\end{split}
\end{equation}
This, in turn, can be used to prove unbiasedness:
\begin{equation}
\begin{split}
\mathbbm{E}[f_{\text{IPS}}(\ldots)] &= \sum_q P(q) \sum_{d:\rho(d \,|\, q_i) > 0}  \zeta_{d, q}\lambda(d \mid \pi_1, \pi_2)
\\
&= \mathbbm{E}[\text{CTR}(\pi_1)] - \mathbbm{E}[\text{CTR}(\pi_2)] = \Delta(\pi_1, \pi_2).
\end{split}
\end{equation}
This proof is only valid under the following requirement:
\begin{equation}
\forall d, q\,  (\zeta_{d, q}\lambda(d \mid \pi_1, \pi_2) > 0 \rightarrow \rho(d \,|\, q) > 0).
\label{eq:requirement}
\end{equation}
In practice, this means that the items in the top-k of either $\pi_1$ or $\pi_2$ need to have a non-zero examination probability under $\pi_0$, i.e., they must have a chance to appear in the top-k under $\pi_0$. 

Besides Requirement~\ref{eq:requirement} the existing counterfactual method~\citep{joachims2017accurately, wang2016learning} is completely indifferent to $\pi_0$ and hence we do not consider it to be an online method.
In the next section, we will introduce an algorithm for choosing and updating $\pi_0$ during logging to minimize the variance of the estimator.
By doing so we turn counterfactual evaluation into an online method.

%% file: sections/04-method.tex

\section{Logging Policy Optimization for Variance Minimization}
Next, we introduce a method aimed at finding a logging policy for the counterfactual estimator that minimizes its variance.

\subsection{Minimizing Variance}

In Section~\ref{sec:related:counterfactual}, we have discussed counterfactual evaluation and established that it is unbiased as long as $\theta$ is known and the logging policy meets Requirement~\ref{eq:requirement}.
The variance of $\Delta_\mathit{IPS}$ depends on the position bias $\theta$, the conditional click probabilities $\zeta$, and the logging policy $\pi_0$.
In contrast with the user-dependent $\theta$ and $\zeta$, the way data is logged by $\pi_0$ is something one can have control over.
The goal of our method is to find the optimal policy that minimizes variance while still meeting Requirement~\ref{eq:requirement}:
\begin{equation}
\pi_0^* = \underset{\pi_0 : \, \pi_0 \text{ meets Req.~\ref{eq:requirement}}}{\argmin} \, \text{Var}\big(\hat{\Delta}_\mathit{IPS}^{\pi_0}\big),
\end{equation}
where $\hat{\Delta}_\mathit{IPS}^{\pi_0}$ is the counterfactual estimator based on data logged using $\pi_0$.

To formulate the variance, we first note that it is an expectation over queries:
\begin{equation}
\text{Var}(\hat{\Delta}) = \sum_q P(q) \text{Var}(\hat{\Delta} \mid q).
\end{equation}
To keep notation short, for the remainder of this section we will write: $\Delta = \Delta(\pi_1, \pi_2)$; $\theta_{d,R} = \theta_\text{rank}(d \mid R)$; $\zeta_{d} = \zeta_{d,q}$; $\lambda_d = \lambda(d \mid \pi_1, \pi_2)$; and $\rho_d = \rho(d \mid q, \pi_0)$. 
Next, we consider the probability of a click pattern $c$, this is simply a possible combination of clicked documents $c(d)=1$ and not-clicked documents $c(d)=0$:
\begin{equation}
\begin{split}
P(c \mid q) &= \sum_{R} \pi_0(R \mid q) \prod_{d : c(d) = 1} \theta_{d,R} \zeta_d \prod_{d : c(d) = 0} (1 - \theta_{d,R} \zeta_d)
\\
&= \sum_{R} \pi_0(R \mid q) P(c \mid R).
\end{split}
\label{eq:clickprob}
\end{equation}
Here, $\pi_0$ has some control over this probability: by deciding the distribution of displayed rankings it can make certain click patterns more or less frequent.
The variance added per query is the squared error of every possible click pattern weighted by the probability of each pattern. 
Let $\sum_c$ sum over every possible click pattern:
\begin{equation}
\text{Var}(\hat{\Delta}_\mathit{IPS}^{\pi_0} \mid q) = \sum_c P(c\mid q) \bigg(\Delta - \sum_{d : c(d) = 1}\frac{\lambda_d}{\rho_d}\bigg)^2.
\label{eq:ipsvariance}
\end{equation}
It is unknown whether there is a closed-form solution for $\pi_0^*$. 
However, the variance function is differentiable.
Taking the derivative reveals a trade-off between two potentially conflicting goals:
\begin{equation}
\begin{split}
\frac{\delta}{\delta \pi_0} \text{Var}(\hat{\Delta}_{IPS}^{\pi_0} \mid q) = \sum_c & \hspace{-1em}
\overbrace{
\left[\frac{\delta}{\delta \pi_0}P(c \mid q)\right] \bigg(\Delta - \sum_{d : c(d) = 1}\frac{\lambda_d}{\rho_d}\bigg)^2
}^{\text{\footnotesize minimize frequency of high-error click patterns}}
\\  + &
\underbrace{
P(c \mid q) \left[\frac{\delta}{\delta \pi_0} \bigg(\Delta - \sum_{d : c(d) = 1}\frac{\lambda_d}{\rho_d}\bigg)^2\right]
}_{\text{\footnotesize minimize error of frequent click patterns}}.
\end{split}
\label{eq:gradient}
\end{equation}
On the one hand, the derivative reduces the frequency of click patterns that result in high error samples, i.e., by updating $\pi_0$ so that these are less likely to occur.
On the other hand, changing $\pi_0$ also affects the propensities $\rho_d$, i.e., if $\pi_0$ makes an item $d$ less likely to be examined, its corresponding value $\lambda_d/\rho_d$ becomes larger, which can lead to a higher error for related click patterns.
The optimal policy has to balance:
\begin{enumerate*}[label=(\roman*)]
\item avoiding showing rankings that lead to high-error click patterns; and
\item avoiding minimizing propensity scores, which increases the errors of corresponding click patterns.
\end{enumerate*}

Our method applies stochastic gradient descent to optimize the logging policy w.r.t.\ the variance.
There are two main difficulties with this approach:
\begin{enumerate*}[label=(\roman*)]
\item the parameters $\theta$ and $\zeta$ are unknown a priori; and
\item the gradients include summations over all possible rankings and all possible click patterns, both of which are computationally infeasible.
\end{enumerate*}
In the following sections, we will detail how \ac{LogOpt} solves both of these problems.

\subsection{Bias \& Relevance Estimation}

In order to compute the gradient in Eq.~\ref{eq:gradient}, the parameters $\theta$ and $\zeta$ have to be known.
\ac{LogOpt} is based on the assumption that accurate estimates of $\theta$ and $\zeta$ suffice to find a near-optimal logging policy.
We note that the counterfactual estimator only requires $\theta$ to be known for unbiasedness~(see Section~\ref{sec:related:counterfactual}).
Our approach is as follows, at given intervals during evaluation we use the available clicks to estimate $\theta$ and $\zeta$.
Then we use the estimated $\hat{\theta}$ to get the current estimate $\hat{\Delta}_\mathit{IPS}(\mathcal{I}, \hat{\theta})$ (Eq.~\ref{eq:ipsestimator}) and optimize w.r.t.\ the estimated variance (Eq.~\ref{eq:ipsvariance}) based on $\hat{\theta}$, $\hat{\zeta}$, and $\hat{\Delta}_\mathit{IPS}(\mathcal{I}, \hat{\theta})$.

For estimating $\theta$ and $\zeta$ we use the existing \ac{EM} approach by \citet{wang2018position}, because it works well in situations where few interactions are available and does not require randomization.
We note that previous work has found randomization-based approaches to be more accurate for estimating $\theta$~\citep{wang2018position, agarwal2019estimating, fang2019intervention}.
However, they require multiple interactions per query and specific types of randomization in their results, by choosing the \ac{EM} approach we do avoid having these requirements.

\subsection{Monte-Carlo-Based Derivatives}
\label{sec:method:derivates}
Both the variance (Eq.~\ref{eq:ipsvariance}) and its gradient (Eq.~\ref{eq:gradient}), include a sum over all possible click patterns.
Moreover, they also include the probability of a specific pattern $P(c \mid q)$ that is based on a sum over all possible rankings (Eq.~\ref{eq:clickprob}).
Clearly, these equations are infeasible to compute under any realistic time constraints.
To solve this issue, we introduce gradient estimation based on Monte-Carlo sampling.
Our approach is similar to that of \citet{ma2020off}, however, we are estimating gradients of variance instead of general performance.

First, we assume that policies place the documents in order of rank and the probability of placing an individual document at rank $x$ only depends on the previously placed documents.
Let $R_{1:x-1}$ indicate the (incomplete) ranking from rank $1$ up to rank $x$, then $\pi_0(d \mid R_{1:x-1}, q)$ indicates the probability that document $d$ is placed at rank $x$ given that the ranking up to $x$ is $R_{1:x-1}$.
The probability of a ranking $R$ up to rank $k$ is thus:
\begin{equation}
\pi_0(R_{1:k} \mid q) = \prod_{x=1}^{k} \pi_0(R_x \mid R_{1:x-1}, q).
\end{equation}
Let $K$ be the length of a complete ranking $R$, the gradient of the probability of a ranking w.r.t.\ a policy is:
\begin{equation}
\frac{\delta \pi_0(R \mid q) }{\delta \pi_0}
 =
\sum_{x=1}^K
  \frac{\pi_0(R \mid q)}{\pi_0(R_x \mid R_{1:x},  q)}
  \left[\frac{\delta \pi_0(R_x \mid R_{1:x-1},  q) }{\delta \pi_0} \right].
\end{equation}
The gradient of the propensity w.r.t.\ the policy (cf.\ Eq.~\ref{eq:prop}) is:
\begin{align}
\frac{\delta \rho(d |\, q)}{\delta \pi_0} &=  \sum_{k=1}^K \theta_k \sum_R \pi_0(R_{1:k-1} |\,  q)   
\Bigg(
\left[ \frac{\delta \pi_0(d |\, R_{1:k-1}, q)}{\delta \pi_0} \right] \nonumber
\\
& \quad   + 
\sum_{x=1}^{k-1} \frac{\pi_0(d |\, R_{1:k-1}, q)}{\pi_0(R_x |\, R_{1:x-1} , q)}\left[ \frac{\delta\pi_0(R_x |\,  R_{1:x-1} , q)}{\delta \pi_0} \right]
\Bigg).
\end{align}
To avoid iterating over all rankings in the $\sum_R$ sum, we sample $M$ rankings:
$R^m \sim \pi_0(R \mid q)$, and a click pattern on each ranking: $c^m \sim P(c \mid R^m)$.
This enables us to make the following approximation:
\begin{align}
\widehat{\rho\text{-grad}}(d)
&=
 \frac{1}{M} \sum_{m=1}^{M} \sum_{k=1}^K \theta_k 
\Bigg(
\left[ \frac{\delta \pi_0(d |\, R^m_{1:k-1}, q)}{\delta \pi_0} \right] 
\\
& \quad \quad \, + 
\sum_{x=1}^{k-1} \frac{\pi_0(d |\, R^m_{1:k-1}, q)}{\pi_0(R^m_x  |\, R^m_{1:x-1} , q)}\left[ \frac{\delta\pi_0(R^m_x  |\, R^m_{1:x-1} , q)}{\delta \pi_0} \right]
\Bigg), \nonumber
\end{align}
since $\frac{\delta \rho(d |\, q)}{\delta \pi_0} \approx \widehat{\rho\text{-grad}}(d, q)$.
In turn, we can use this to approximate the second part of Eq.~\ref{eq:gradient}:
\begin{equation}
\widehat{\text{error-grad}}(c) =
2\bigg(\Delta - \sum_{d : c(d) = 1}\frac{\lambda_d}{\rho_d}\bigg) \sum_{d : c(d) = 1}\frac{\lambda_d}{\rho_d^2} \widehat{\rho\text{-grad}}(d),
\end{equation}
we approximate the first part of Eq.~\ref{eq:gradient} with:
\begin{align}
&\widehat{\text{freq-grad}}(R, c) = \\
& \bigg(\Delta - \sum_{d : c(d) = 1}\frac{\lambda_d}{\rho_d}\bigg)^2
 \sum_{x=1}^K
\frac{1}{\pi_0(R_x \mid R_{1:x-1},  q)}
\left[\frac{\delta \pi_0(R_x \mid R_{1:x-1},  q) }{\delta \pi_0} \right].
\nonumber
\end{align}
Together, they approximate the complete gradient (cf.\ Eq.~\ref{eq:gradient}):
\begin{equation}
\begin{split}
&\frac{\delta \text{Var}(\hat{\Delta}_{IPS}^{\pi_0} \mid q)}{\delta \pi_0} 
\approx {}\\
&\mbox{}\quad
\frac{1}{M}\sum_{m=1}^M \widehat{\text{freq-grad}}(R^m, c^m) + \widehat{\text{error-grad}}(c^m).
\label{eq:approxgrad}
\end{split}
\end{equation}
Therefore, we can approximate the gradient of the variance w.r.t.\ a logging policy $\pi_0$, based on rankings sampled from $\pi_0$ and our current estimated click model $\hat{\theta}$, $\hat{\zeta}$, while staying computationally feasible.\footnote{For a more detailed description see Appendix~B in the supplementary material.}

\subsection{Summary}

We have summarized the \ac{LogOpt} method in Algorithm~\ref{alg:logopt}.
The algorithm requires a set of historical interactions $\mathcal{I}$ and two rankers $\pi_1$ and $\pi_2$ to compare.
Then by fitting a click model on $\mathcal{I}$ using an \ac{EM}-procedure (Line~\ref{line:clickmodel}) an estimate of observation bias $\hat{\theta}$ and document relevance $\hat{\zeta}$ is obtained.
Using $\hat{\theta}$, an estimate of the difference in observation probabilities $\hat{\lambda}$ is computed (Line~\ref{line:lambda} and cf.\ Eq~\ref{eq:lambda}), and an estimate of the CTR difference $\hat{\Delta}(\pi_1, \pi_2)$ (Line~\ref{line:delta} and cf.\ Eq~\ref{eq:ipsestimator}).
Then the optimization of a new logging policy $\pi_0$ begins:
A query is sampled from $\mathcal{I}$ (Line~\ref{line:query}), and for that query $M$ rankings are sampled from the current $\pi_0$ (Line~\ref{line:rankings}), then for each ranking a click pattern is sampled using $\hat{\theta}$ and $\hat{\zeta}$ (Line~\ref{line:clicks}).
Finally, using the sampled rankings and clicks, $\hat{\theta}$, $\hat{\lambda}$, and $\hat{\Delta}(\pi_1, \pi_2)$, the gradient is now approximated using Eq.~\ref{eq:approxgrad} (Line~\ref{line:grad}) and the policy $\pi_0$ is updated accordingly (Line~\ref{line:update}).
This process can be repeated for a fixed number of steps, or until the policy has converged.

This concludes our introduction of \ac{LogOpt}: the first method that optimizes the logging policy for faster convergence in counterfactual evaluation.
We argue that \ac{LogOpt} turns counterfactual evaluation into online evaluation, because it instructs which rankings should be displayed for the most efficient evaluation.
The ability to make interventions like this is the defining characteristic of an online evaluation method.

\begin{algorithm}[t]
\caption{\acf{LogOpt}} 
\label{alg:logopt}
\begin{algorithmic}[1]
\STATE \textbf{Input}: Historical interactions: $\mathcal{I}$; rankers to compare $\pi_1, \pi_2$.
\STATE $\hat{\theta}, \hat{\zeta} \leftarrow \text{infer\_click\_model}(\mathcal{I})$ \hfill \textit{\small // estimate bias using EM} \label{line:clickmodel}
\STATE $\hat{\lambda} \leftarrow \text{estimated\_observance}(\hat{\theta}, \pi_1, \pi_2)$ \hfill \textit{\small // estimate $\lambda$ cf.\ Eq~\ref{eq:lambda}} \label{line:lambda}
\STATE $\hat{\Delta}(\pi_1, \pi_2) \leftarrow \text{estimated\_CTR}(\mathcal{I}, \hat{\lambda}, \hat{\theta})$ \hfill \textit{\small // CTR diff. cf.\ Eq~\ref{eq:ipsestimator}} \label{line:delta}
\STATE $\pi_0 \leftarrow \text{init\_policy}()$ \hfill \textit{\small // initialize logging policy} 
\FOR{$j \in \{1,2,\ldots\}$}
\STATE $q \sim P(q\mid \mathcal{I})$ \hfill \textit{\small // sample a query from interactions} \label{line:query}
\STATE $\mathcal{R} \leftarrow \{R^1, R^2, \ldots, R^M\} \sim \pi_0(R \mid q)$ \hfill \textit{\small // sample $M$ rankings} \label{line:rankings}
\STATE $\mathcal{C} \leftarrow \{c^1, c^2, \ldots, c^M\} \sim P(c \mid \mathcal{R})$ \hfill \textit{\small // sample $M$ click patterns}  \label{line:clicks}
\STATE $\hat{\delta} \leftarrow \text{approx\_grad}(\mathcal{R}, \mathcal{C}, \hat{\lambda}, \hat{\theta}, \hat{\Delta}(\pi_1, \pi_2))$  \label{line:grad}
\hfill \textit{\small // using Eq.~\ref{eq:approxgrad}}
\STATE $\pi_0 \leftarrow \text{update}(\pi_0, \hat{\delta})$  \label{line:update}
\hfill \textit{\small // update using approx. gradient}
\ENDFOR
\RETURN $\pi_0$ 
\end{algorithmic}
\end{algorithm}

%% file: sections/05-experimental-setup.tex
\section{Experimental Setup}

We ran semi-synthetic experiments that are prevalent in online and counterfactual evaluation~\citep{oosterhuis2020topkrankings, hofmann2011probabilistic, joachims2017unbiased}.
User-issued queries are simulated by sampling from learning to rank datasets; each dataset contains a preselected set of documents per query.
We use Yahoo! Webscope~\citep{Chapelle2011} and MSLR-WEB30k~\citep{qin2013introducing}; they both contain 5-grade relevance judgements for all preselected query-document pairs.
For each sampled query, we let the evaluation method decide which ranking to display and then simulate clicks on them using probabilistic click models.

To simulate position bias, we use the rank-based probabilities of \citet{joachims2017unbiased}:
\begin{equation}
P(o(d) = 1 \mid R, q) = \frac{1}{\text{rank}(d \mid R)}.
\end{equation}
If observed, the click probability is determined by the relevance label of the dataset (ranging from 0 to 4).
More relevant items are more likely to be clicked, yet non-relevant documents still have a non-zero click probability:
\begin{equation}
P(c(d) = 1 \mid o(d) = 1, q) =  0.225 \cdot \text{relevance\_label}(q,d) + 0.1.
\end{equation}
Spread over both datasets, we generated {2,000} rankers and created {1,000} ranker-pairs.
We aimed to generate rankers that are likely to be compared in real-world scenarios; unfortunately, no simple distribution of such rankers is available.
Therefore, we tried to generate rankers that have (at least) a decent \ac{CTR} and that span a variety of ranking behaviors.
Each ranker was optimized using LambdaLoss~\citep{wang2018lambdaloss} based on the labelled data of 100 sampled queries; each ranker is based on a linear model that only uses a random sample of 50\% of the dataset features.
Figure~\ref{fig:ctrdistribution} displays the resulting \ac{CTR} distribution; it appears to follow a normal distribution.

For each ranker-pair and method, we sample $3 \cdot 10^6$ queries and calculate their \ac{CTR} estimates for different numbers of queries.
We considered three metrics:
\begin{enumerate*}[label=(\roman*)]
\item The binary error: whether the estimate correctly predicts which ranker should be preferred.
\item The absolute error: the absolute difference between the estimate and the true $\mathbbm{E}[\text{CTR}]$ difference:
\end{enumerate*}
\begin{equation}
\text{absolute-error} = |\Delta(\pi_1, \pi_2) - \hat{\Delta}(\mathcal{I})|.
\end{equation}
And
\begin{enumerate*}[label=(\roman*),resume]
\item the mean squared error: the squared error \emph{per sample} (not the final estimate); if the estimator is unbiased this is equivalent to the variance:
\end{enumerate*}
\begin{equation}
\text{mean-squared-error} = \frac{1}{N}\sum_{i=1}^N (\Delta(\pi_1, \pi_2) - x_i)^2.
\end{equation}
We compare \ac{LogOpt} with the following baselines:
\begin{enumerate*}[label=(\roman*)]
\item A/B testing (with equal probabilities for each ranker),
\item Team-Draft Interleaving,
\item Probabilistic Interleaving (with $\tau = 4$), and
\item Optimized Interleaving (with the inverse rank scoring function).
\end{enumerate*}
Furthermore, we compare \ac{LogOpt} with other choices of logging policies:
\begin{enumerate*}[label=(\roman*)]
\item uniform sampling,
\item A/B testing: showing either the ranking of A or B with equal probability, and
\item an Oracle logging policy: applying \ac{LogOpt} to the true relevances $\zeta$ and position bias $\theta$.
\end{enumerate*}
We also consider \ac{LogOpt} both in the case where $\theta$ is known \emph{a priori}, or where it has to be estimated still.
Because estimating $\theta$ and optimizing the logging policy $\pi_0$ is time-consuming, we only update $\hat{\theta}$ and $\pi_0$ after $10^3$, $10^4$, $10^5$ and $10^6$ queries.
The policy \ac{LogOpt} optimizes uses a neural network with 2 hidden layers consisting of 32 units each.
The network computes a score for every document, then a softmax is applied to the scores to create a distribution over documents.

{\renewcommand{\arraystretch}{0.93}
\begin{figure}[t]
\centering
\begin{tabular}{l l}
 \multicolumn{1}{c}{ \footnotesize \hspace{1.8em} Yahoo Webscope}
&
 \multicolumn{1}{c}{ \footnotesize \hspace{1.8em} MSLR Web30k}
\\
\includegraphics[scale=0.3]{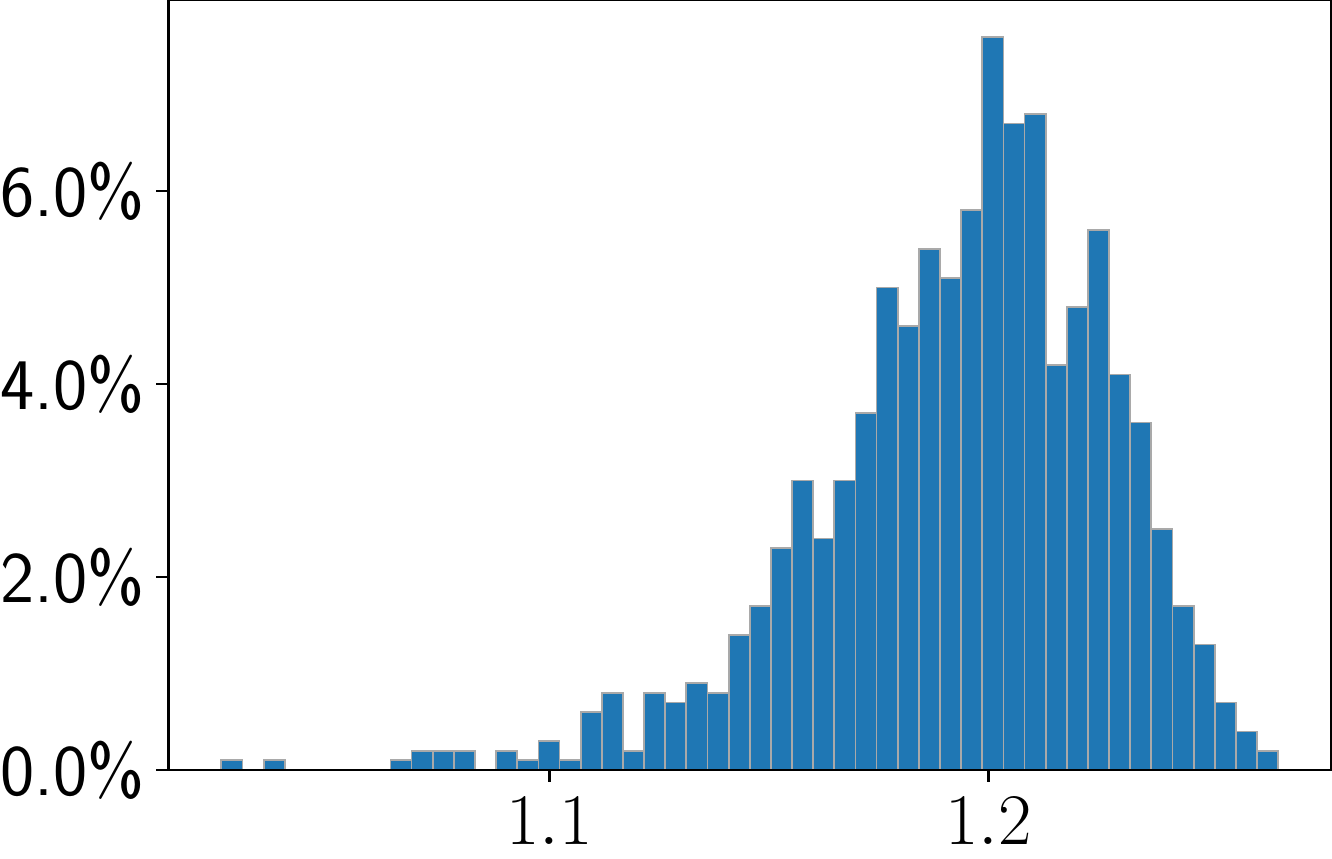} &
\includegraphics[scale=0.3]{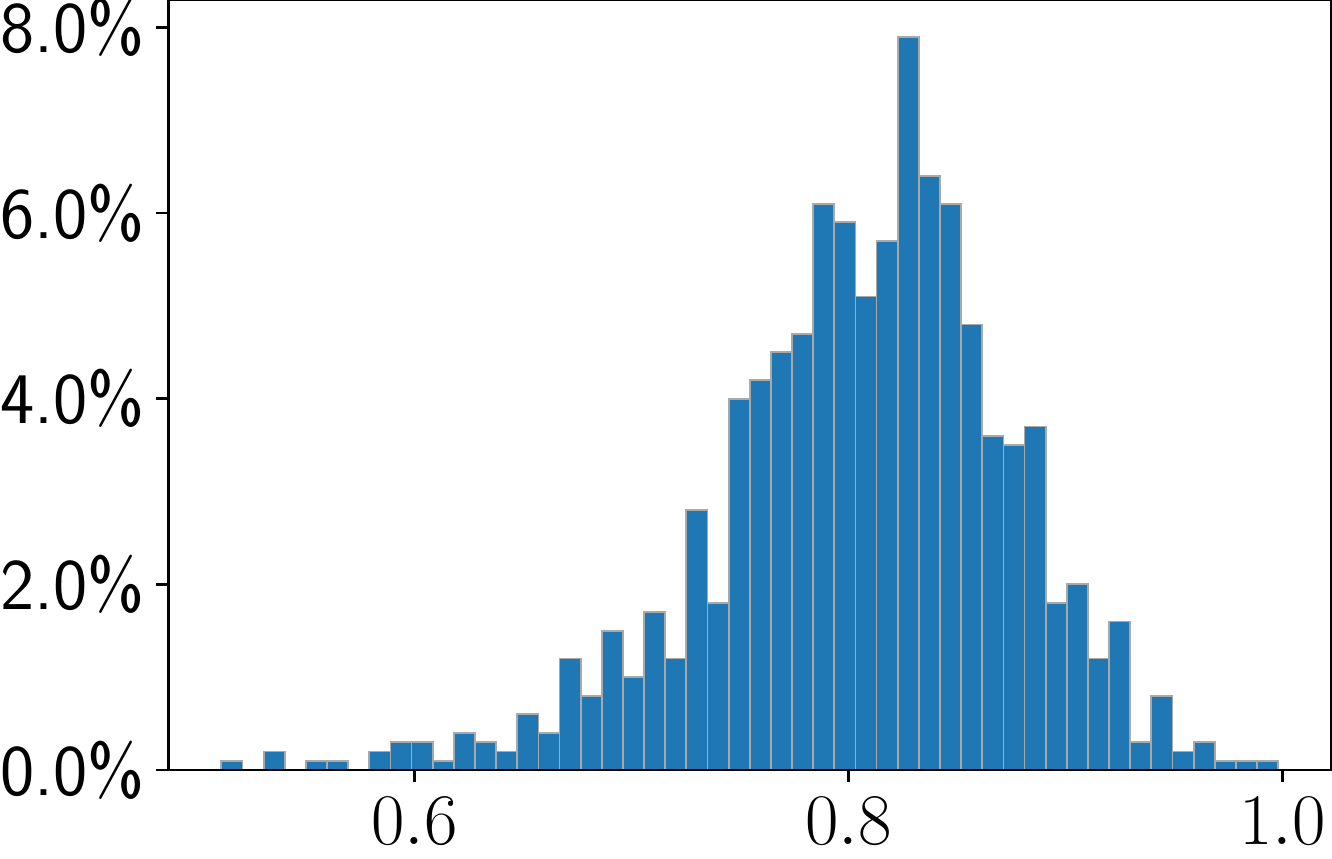} 
\end{tabular}
\caption{
The \ac{CTR} distribution of the 2000 generated rankers, 1000 were generated per dataset.
}
\label{fig:ctrdistribution}
\end{figure}
}

%% file: sections/06-results.tex
\pdfoutput=1

{\renewcommand{\arraystretch}{0.1}
\begin{figure*}[t]
\centering
\begin{tabular}{c r r r }
&
 \multicolumn{1}{c}{\footnotesize  Binary Error}
&
 \multicolumn{1}{c}{\footnotesize Absolute Error}
&
 \multicolumn{1}{c}{\footnotesize Mean Squared Error }

\\
\rotatebox[origin=lt]{90}{\hspace{1.4em} \footnotesize Yahoo Webscope} &
\includegraphics[scale=0.3]{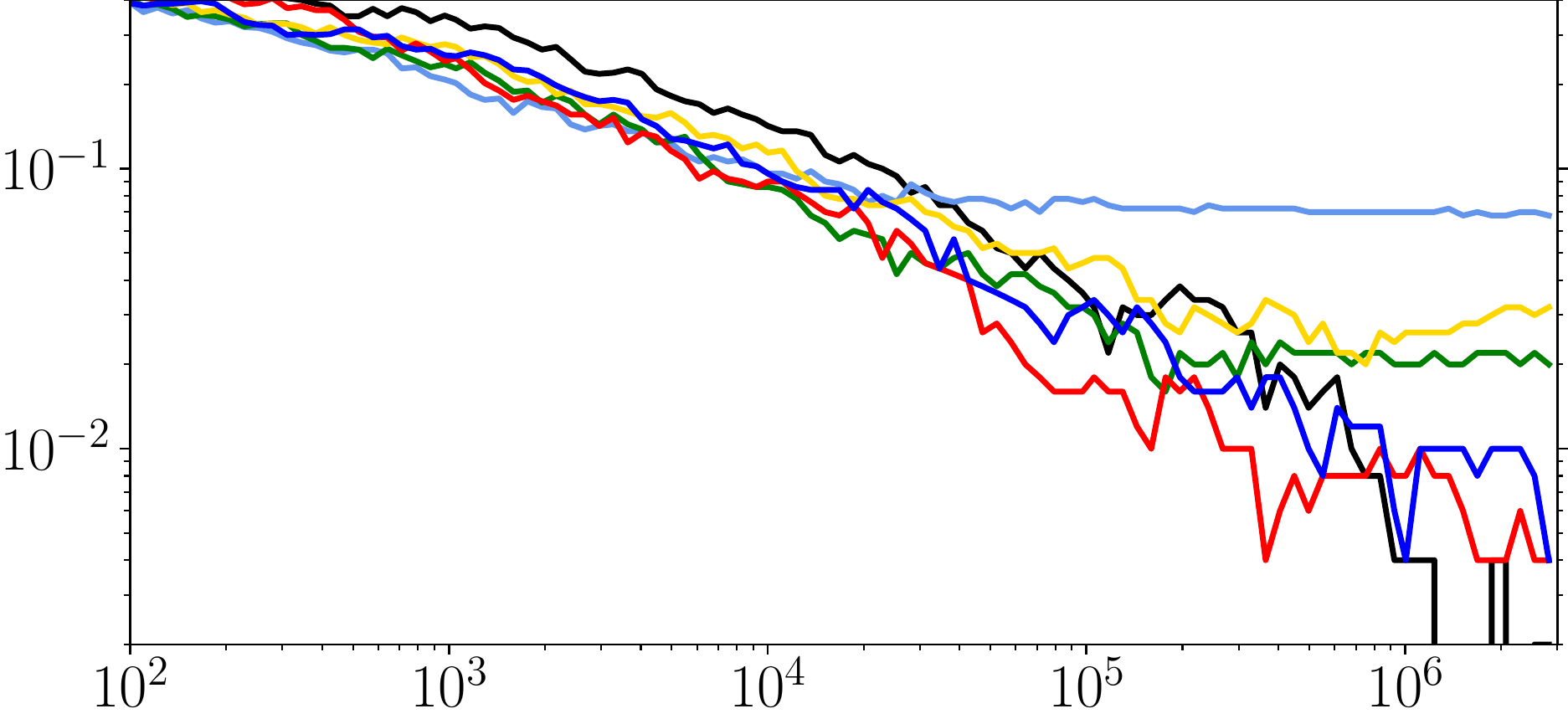} &
\includegraphics[scale=0.3]{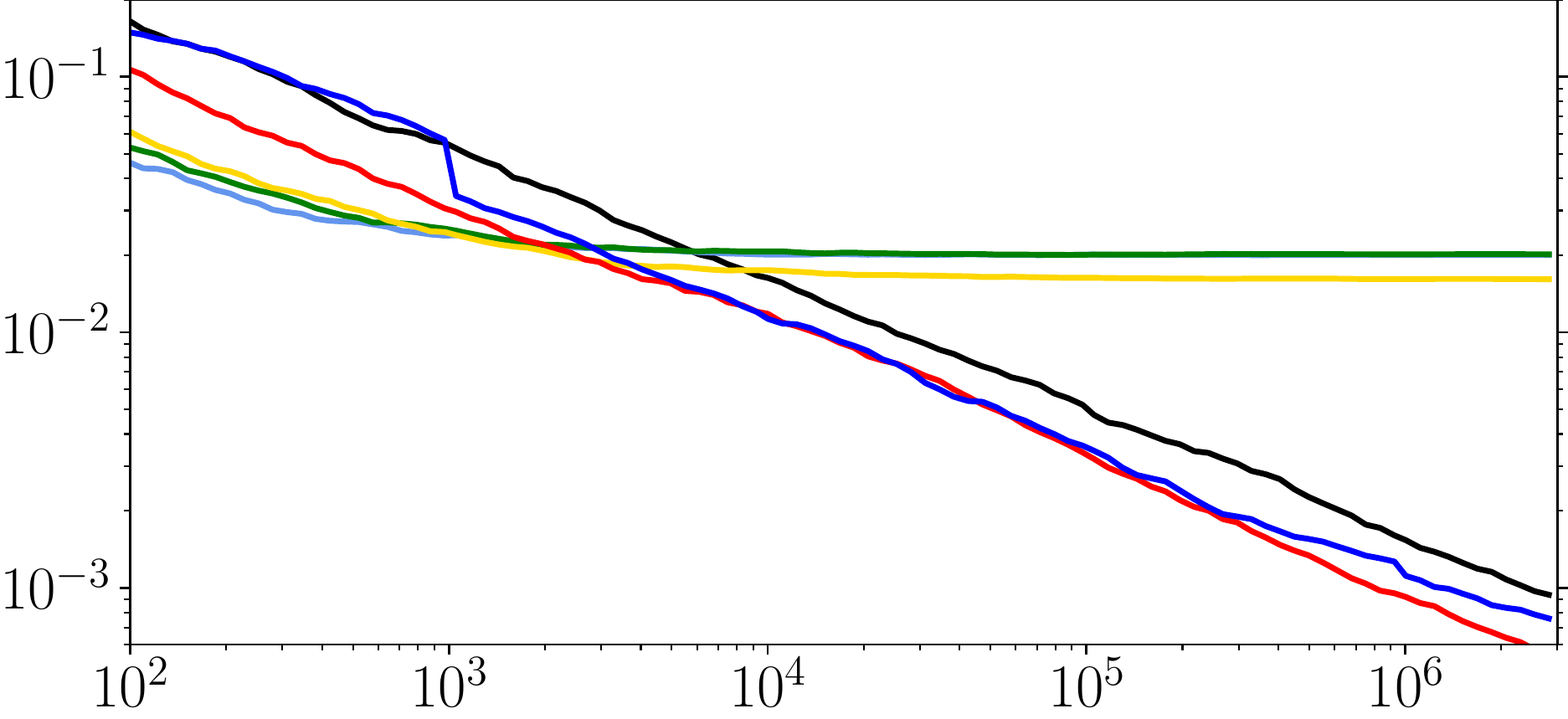} &
\includegraphics[scale=0.3]{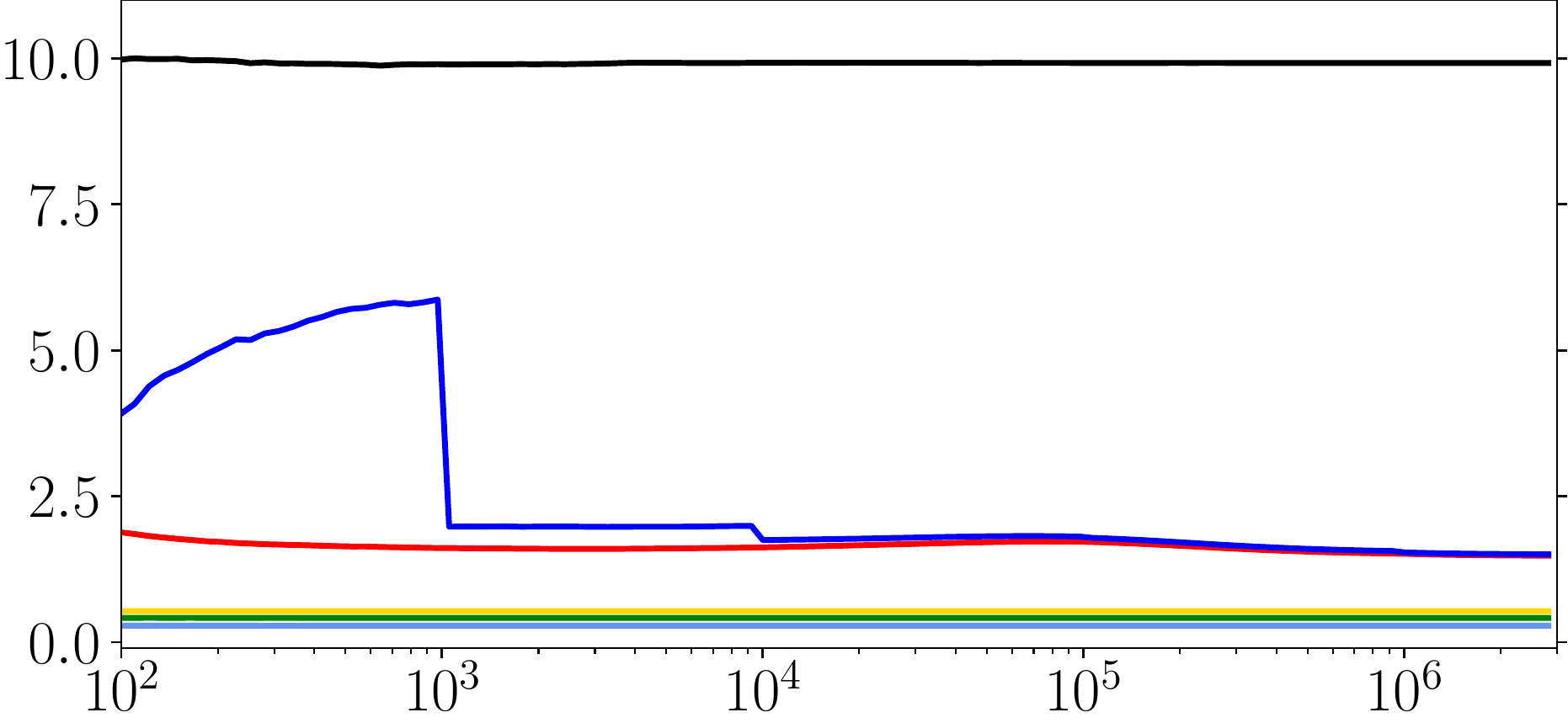} 
\\
\rotatebox[origin=lt]{90}{\hspace{1.5em} \footnotesize MSLR Web30k} &
\includegraphics[scale=0.3]{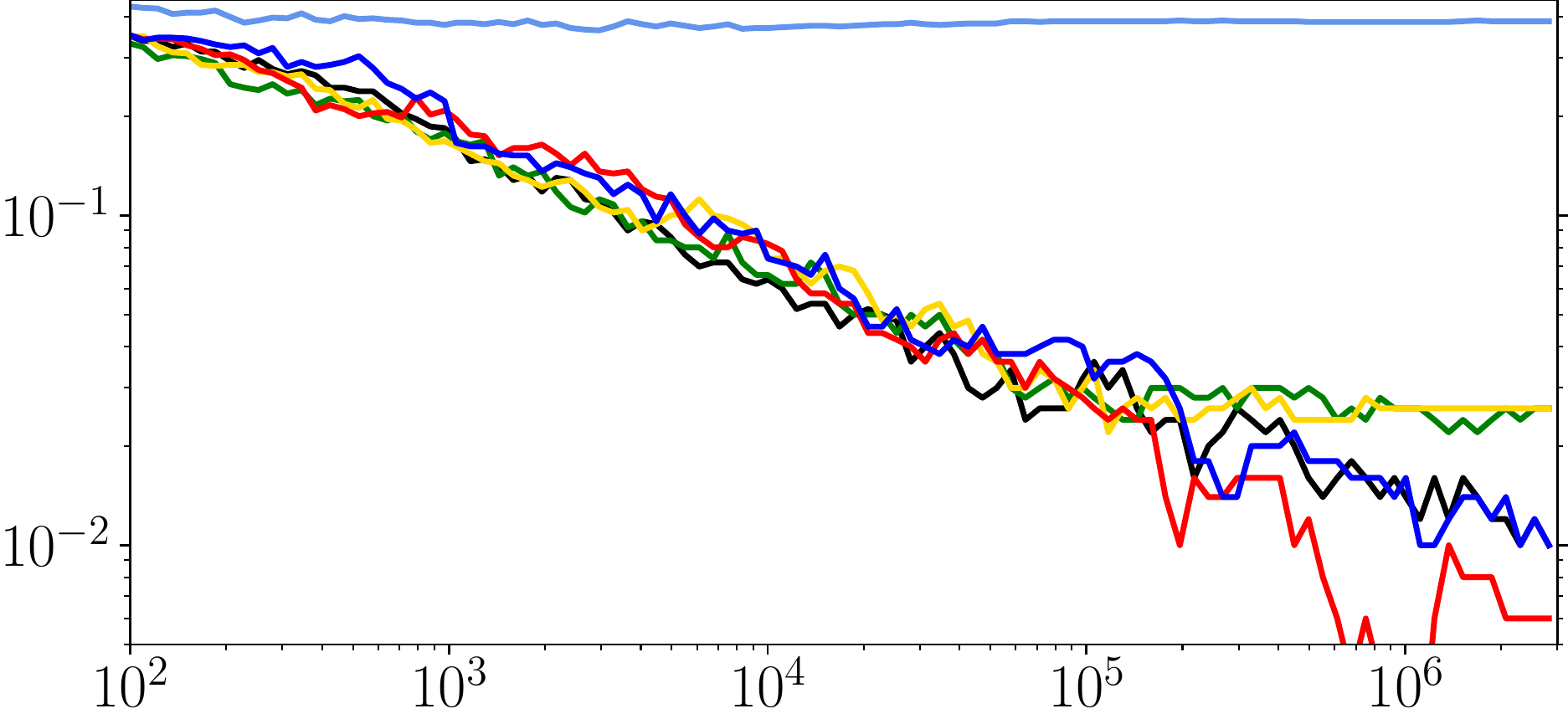} &
\includegraphics[scale=0.3]{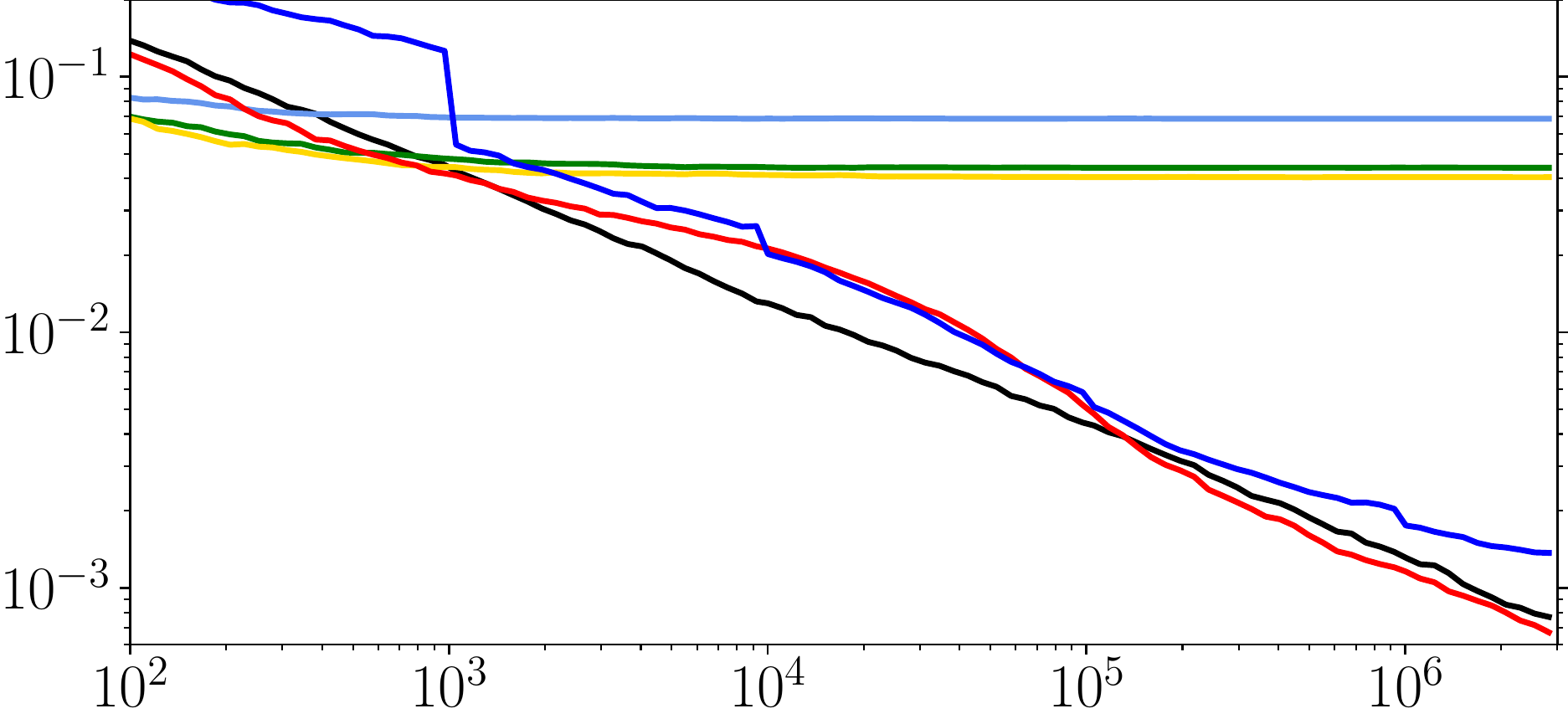} &
\includegraphics[scale=0.3]{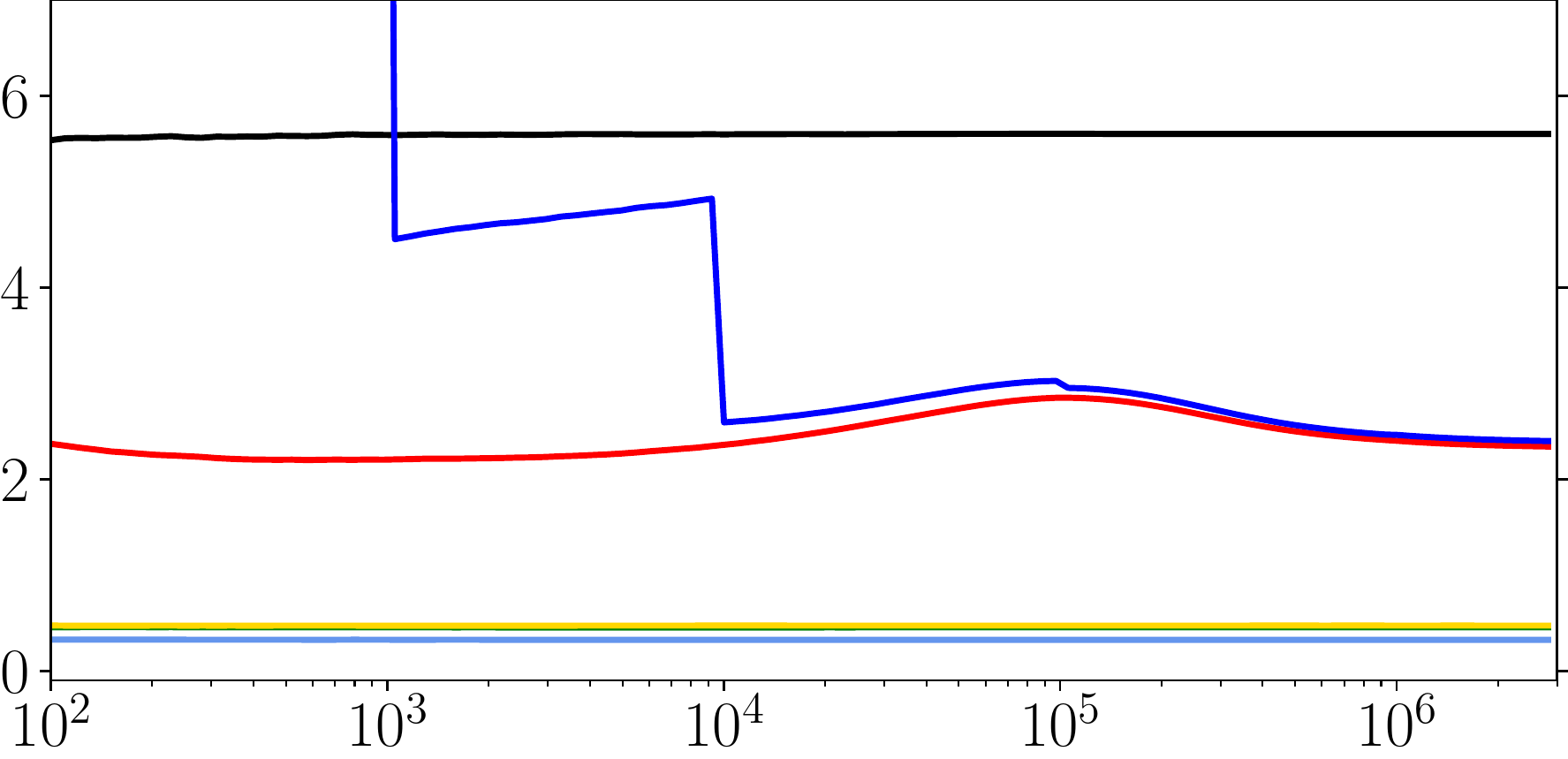} 
\\
& \multicolumn{1}{c}{\footnotesize \hspace{0.5em} Number of Queries Issued}
& \multicolumn{1}{c}{\footnotesize \hspace{0.5em} Number of Queries Issued}
& \multicolumn{1}{c}{\footnotesize \hspace{0.5em} Number of Queries Issued}
 \vspace{0.5em}
\\
 \multicolumn{4}{c}{
 \includegraphics[scale=.4]{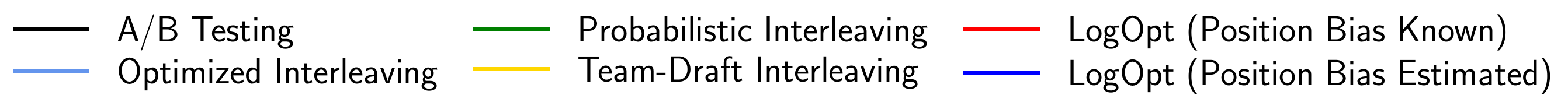}
} 
\end{tabular}
\caption{
Comparison of \ac{LogOpt} with other online methods; displayed results are an average over 500 comparisons.
}
\label{fig:baselines}
\end{figure*}
}

{\renewcommand{\arraystretch}{0.1}
\begin{figure*}[t]
\centering
\begin{tabular}{c r r r }
&
 \multicolumn{1}{c}{\footnotesize Binary Error}
&
 \multicolumn{1}{c}{\footnotesize Absolute Error}
&
 \multicolumn{1}{c}{\footnotesize Mean Squared Error }

\\
\rotatebox[origin=lt]{90}{\hspace{1.4em} \footnotesize Yahoo Webscope} &
\includegraphics[scale=0.3]{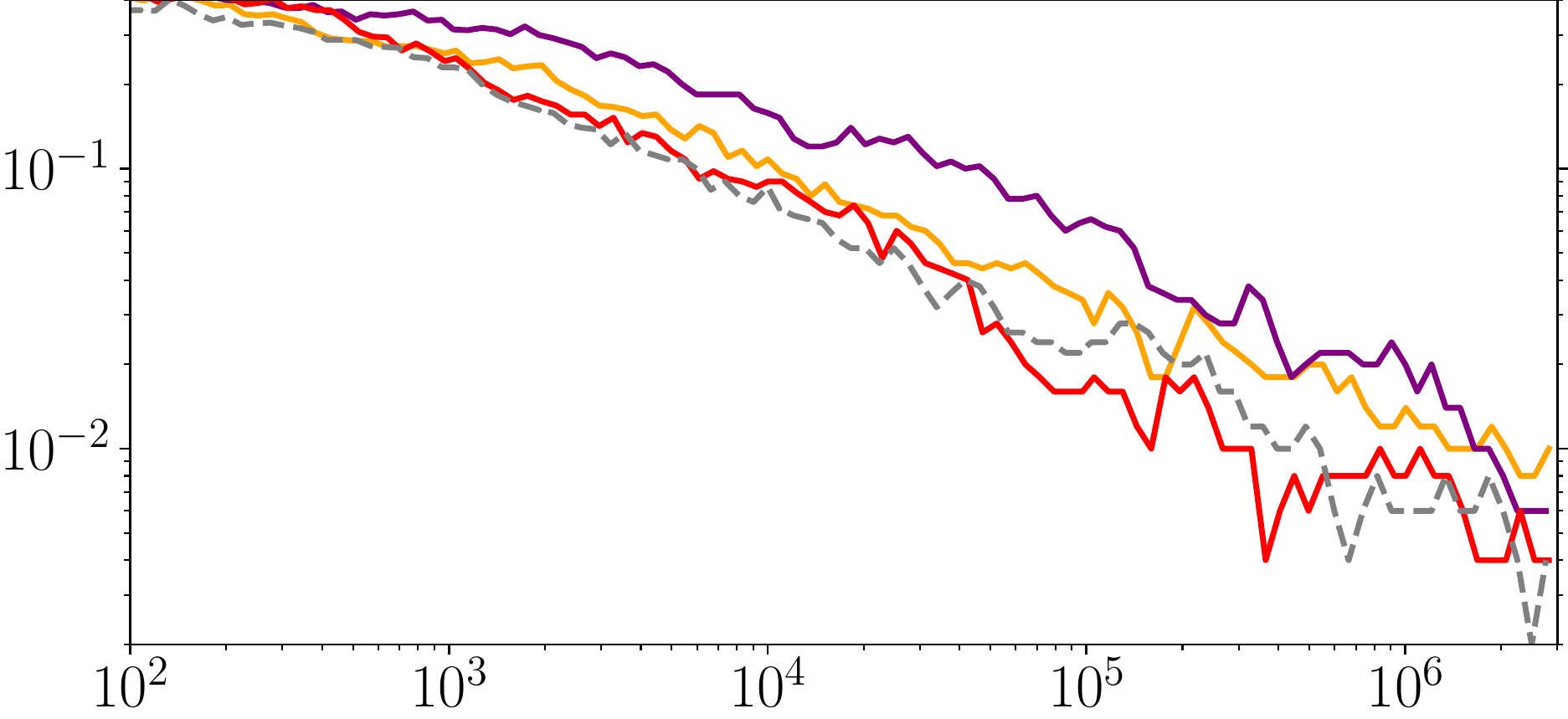} &
\includegraphics[scale=0.3]{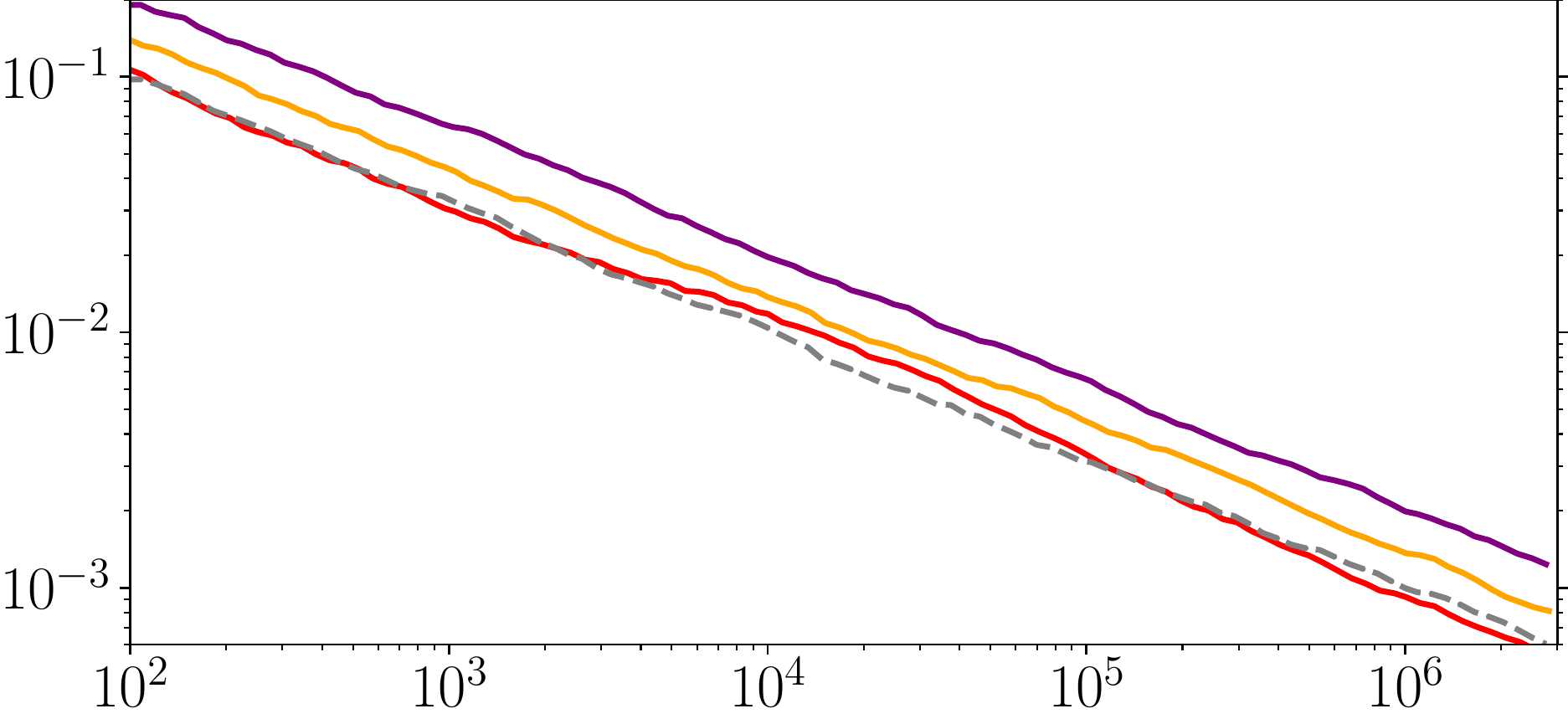} &
\includegraphics[scale=0.3]{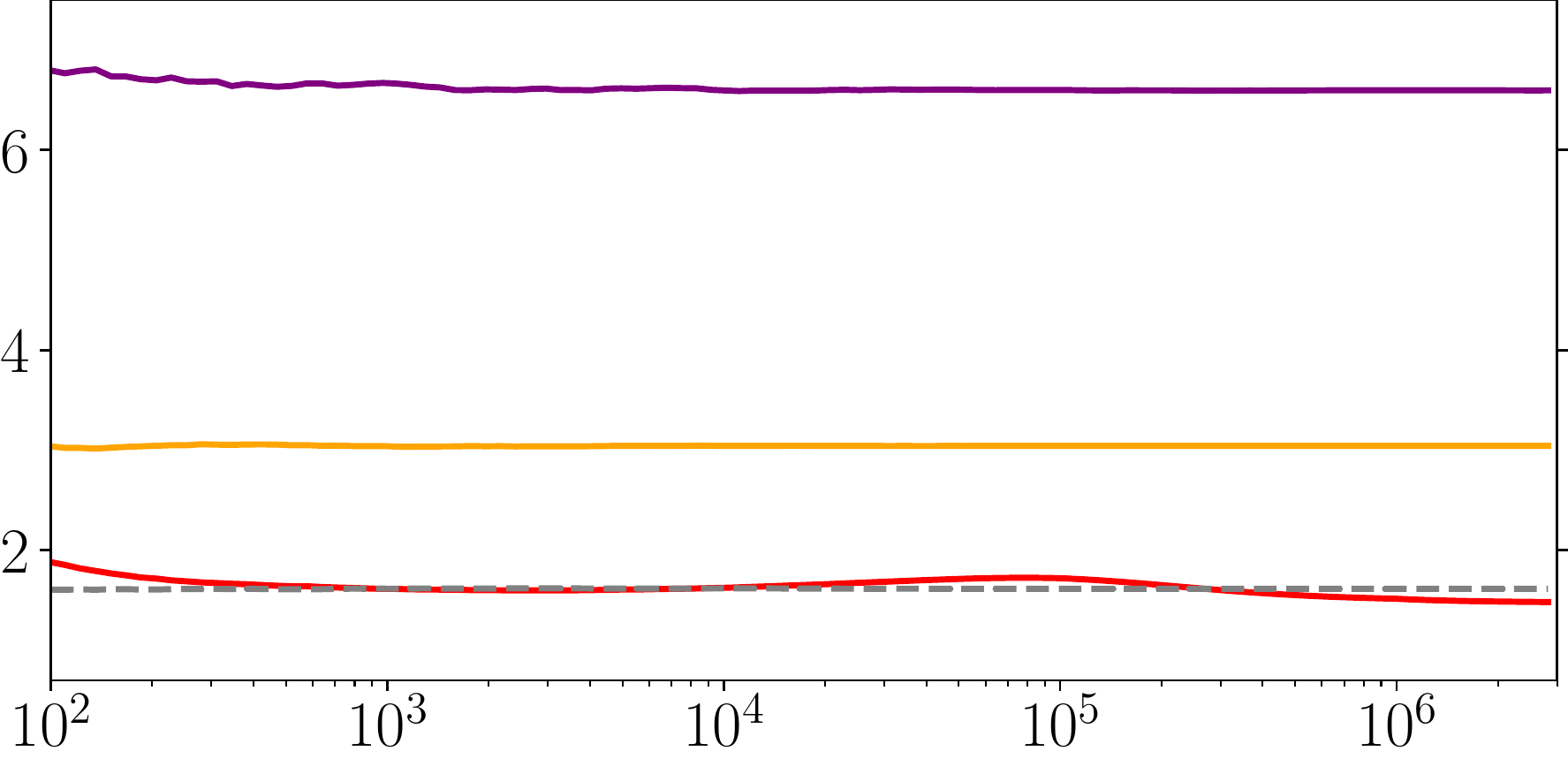} 
\\
\rotatebox[origin=lt]{90}{\hspace{1.5em} \footnotesize MSLR Web30k} &
\includegraphics[scale=0.3]{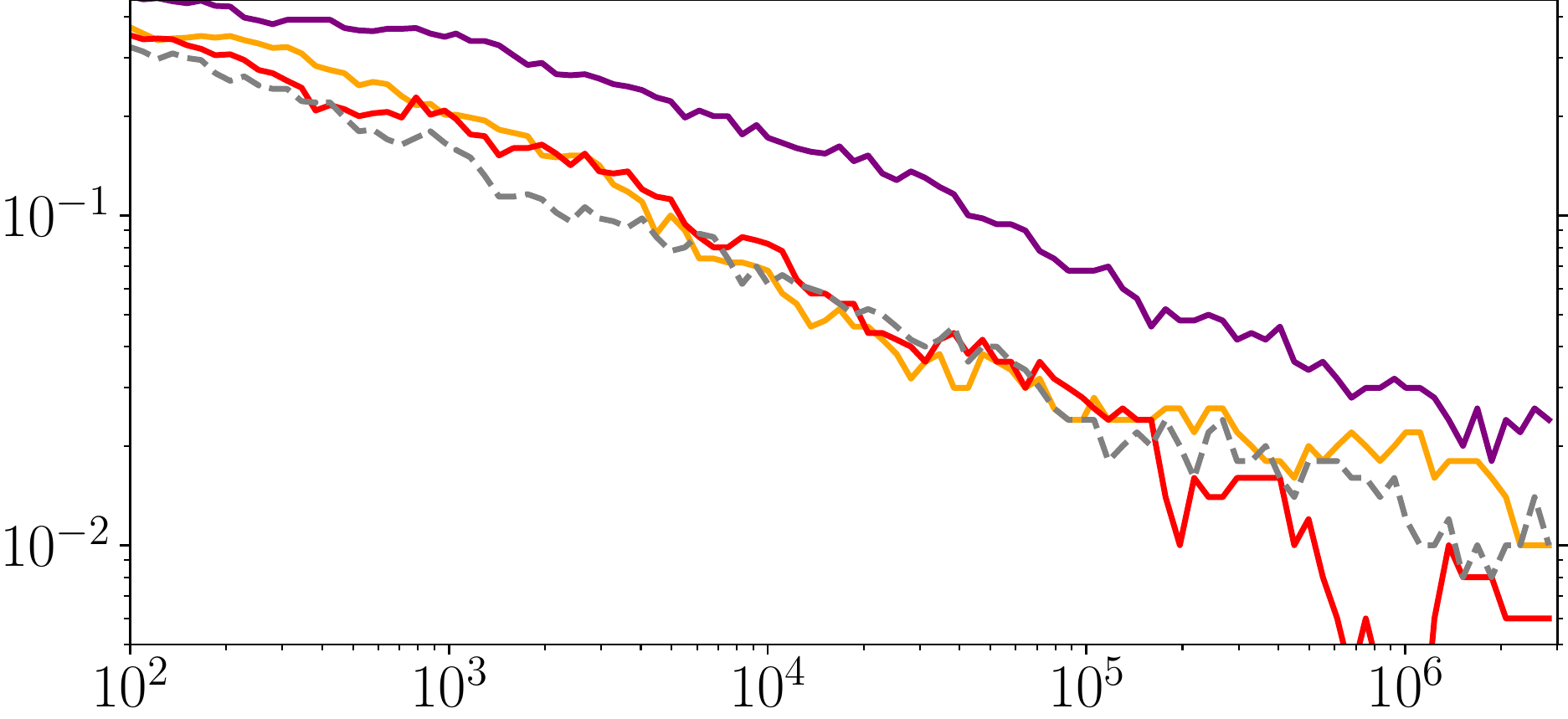} &
\includegraphics[scale=0.3]{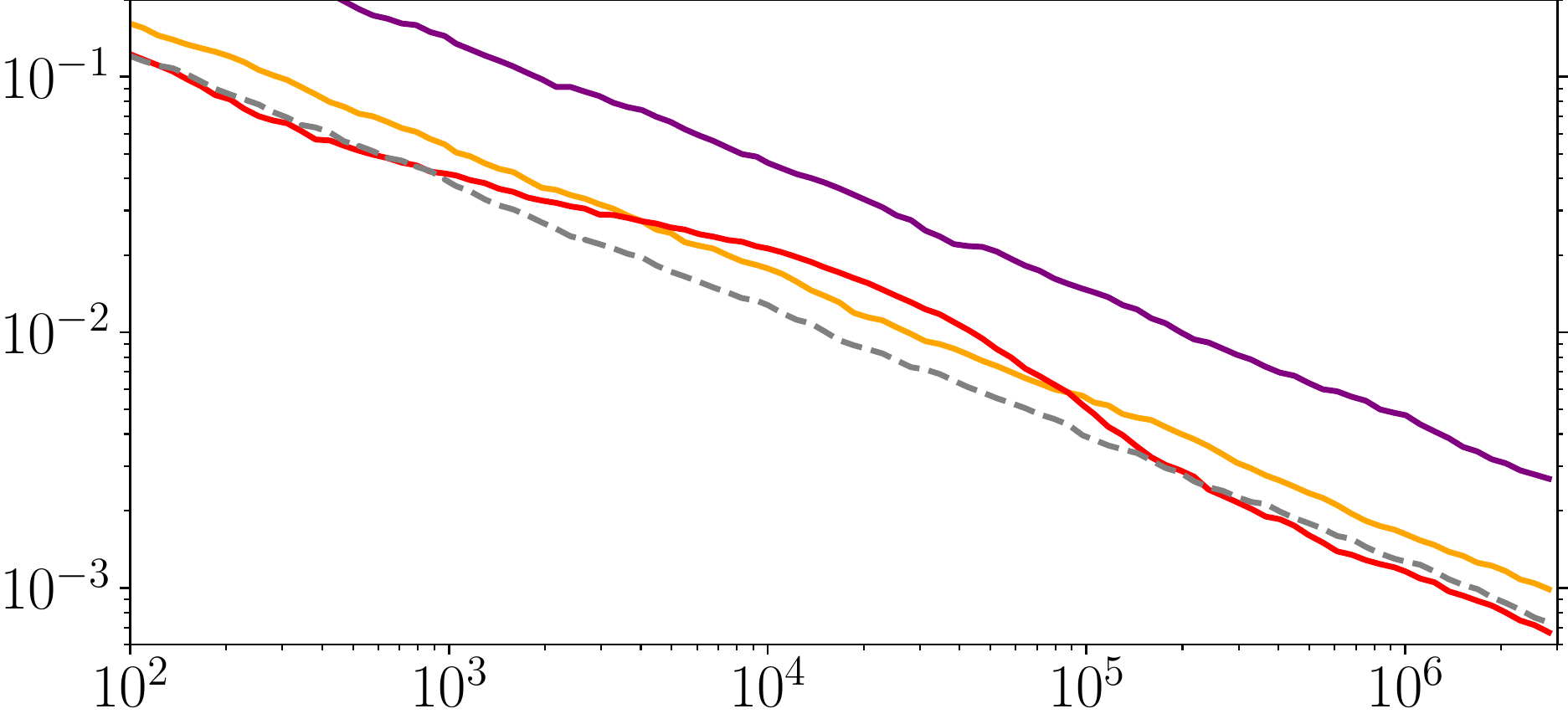} &
\includegraphics[scale=0.3]{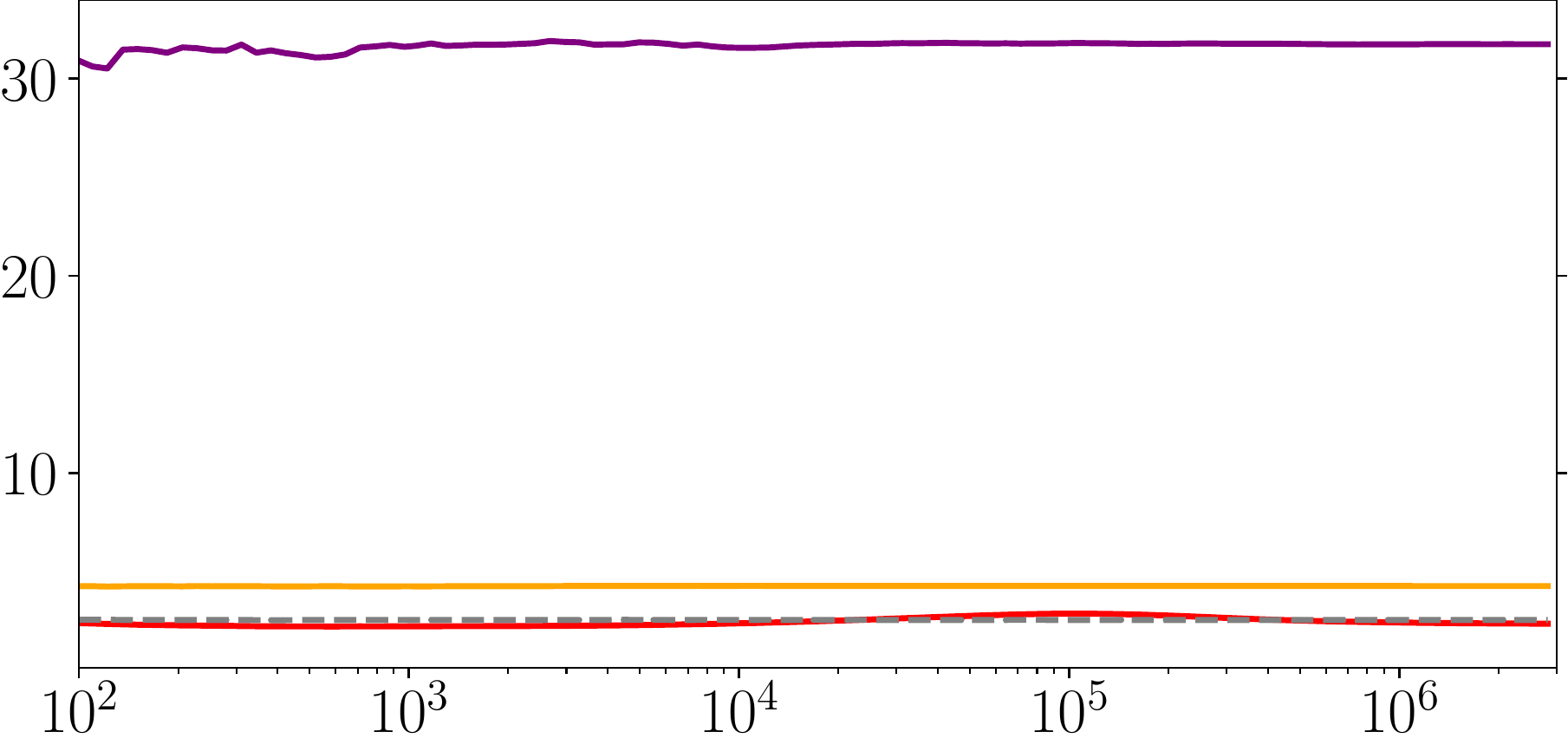} 
\\
& \multicolumn{1}{c}{\footnotesize \hspace{0.5em} Number of Queries Issued}
& \multicolumn{1}{c}{\footnotesize \hspace{0.5em} Number of Queries Issued}
& \multicolumn{1}{c}{\footnotesize \hspace{0.5em} Number of Queries Issued}
 \vspace{0.5em}
\\
 \multicolumn{4}{c}{
 \includegraphics[scale=.4]{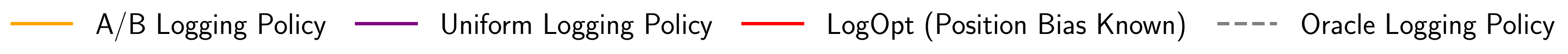}
} 
\end{tabular}
\caption{
Comparison of logging policies for counterfactual evaluation; displayed results are an average over 500 comparisons.
}
\label{fig:loggingpolicy}
\end{figure*}
}

\section{Results}
Our results are displayed in Figures~\ref{fig:baselines},~\ref{fig:loggingpolicy},~and~\ref{fig:errordistribution}.
Figure~\ref{fig:baselines} shows the results comparing \ac{LogOpt} with other online evaluation methods;
Figure~\ref{fig:loggingpolicy} compares \ac{LogOpt} with counterfactual evaluation using other logging policies;
and finally, Figure~\ref{fig:errordistribution} shows the distribution of binary errors for each method after $3 \cdot 10^6$ sampled queries.

\subsection{Performance of \ac{LogOpt}}

In Figure~\ref{fig:baselines} we see that, unlike interleaving methods, counterfactual evaluation with \ac{LogOpt} continues to decrease both its binary error and its absolute error as the number of queries increases.
While interleaving methods converge at a binary error of at least 2.2\% and an absolute error greater than $0.01$, \ac{LogOpt} appears to converge towards zero errors for both.
This is expected as \ac{LogOpt} is proven to be unbiased when the position bias is known. Interestingly, we see similar behavior from \ac{LogOpt} with estimated position bias.
Both when bias is known or estimated, \ac{LogOpt} has a lower error than the interleaving methods after $2 \cdot 10^3$ queries.
Thus we conclude that interleaving methods converge faster and have an initial period where their error is lower, but are biased.
In contrast, by being unbiased, \ac{LogOpt} converges on a lower error eventually. 

If we use Figure~\ref{fig:baselines} to compare \ac{LogOpt} with A/B testing, we see that on both datasets \ac{LogOpt} has a considerably smaller mean squared error.
Since both methods are unbiased, this means that \ac{LogOpt} has a much lower variance and thus is expected to converge faster.
On the Yahoo dataset we observe this behavior, both in terms of binary error and absolute error and regardless of whether the bias is estimated, \ac{LogOpt} requires half as much data as A/B testing to reach the same level or error.
Thus, on Yahoo \ac{LogOpt} is roughly twice as data-efficient than A/B testing.
On the MSLR dataset it is less clear whether \ac{LogOpt} is noticeably more efficient: after $10^4$ queries the absolute error of \ac{LogOpt} is twice as high, but after $10^5$ queries it has a lower error than A/B testing.
We suspect that the relative drop in performance around $10^4$ queries is due to \ac{LogOpt} overfitting on incorrect $\hat{\zeta}$ values, however, we were unable to confirm this.
Hence, \ac{LogOpt} is just as efficient as, or even more efficient than, A/B testing, depending on the circumstances.

Finally, when we use Figure~\ref{fig:loggingpolicy} to compare \ac{LogOpt} with other logging policy choices, we see that \ac{LogOpt} mostly approximates the optimal Oracle logging policy.
In contrast, the uniform logging policy is very data-inefficient on both datasets it requires around ten times the number of queries to reach the same level or error as \ac{LogOpt}.
The A/B logging policy is a better choice than the uniform logging policy, but apart from the dip in performance on the MSLR dataset, it appears to require twice as many queries as \ac{LogOpt}.
Interestingly, the performance of \ac{LogOpt} is already near the Oracle when only $10^2$ queries have been issued.
With such a small number of interactions, accurately estimating the relevances $\zeta$ should not be possible, thus it appears that in order for \ac{LogOpt} to find an efficient logging policy the relevances $\zeta$ are not important.
This must mean that only the differences in behavior between the rankers (i.e.\ $\lambda$) have to be known for \ac{LogOpt} to be efficient.
Overall, these results show that \ac{LogOpt} can greatly increase the efficiency of counterfactual estimation.

\subsection{Bias of Interleaving}
\label{sec:biasinterleaving}
Our results in Figure~\ref{fig:baselines} clearly illustrate the bias of interleaving methods: each of them systematically infers incorrect preferences in (at least) 2.2\% of the ranker-pairs.
These errors are systematic since increasing the number of queries from $10^5$ to $3\cdot10^6$ does not remove any of them.
Additionally, the combination of the lowest mean-squared-error with a worse absolute error than A/B testing after $10^4$ queries, indicates that interleaving results in a low variance at the cost of bias.
To better understand when these systematic errors occur, we show the distribution of binary errors w.r.t.\ the \ac{CTR} differences of the associated ranker-pairs in Figure~\ref{fig:errordistribution}.
Here we see that most errors occur on ranker-pairs where the \ac{CTR} difference is smaller than 1\%, and that of all comparisons the percentage of errors greatly increases as the \ac{CTR} difference decreases below 1\%.
This suggests that interleaving methods are unreliable to detect preferences when differences are 1\% \ac{CTR} or less.

It is hard to judge the impact this bias may have in practice.
On the one hand, a 1\% \ac{CTR} difference is far from negligible: generally a 1\% increase in \ac{CTR} is considered an impactful improvement in the industry.
On the other hand, our results are based on a single click model with specific values for position bias and conditional click probabilities.
While our results strongly prove interleaving is biased, we should be careful not to generalize the size of the observed systematic error to all other ranking settings.

Previous work has performed empirical studies to evaluate various interleaving methods with real users.
\citet{chapelle2012large} applied interleaving methods to compare ranking systems for three different search engines, and found team-draft interleaving to highly correlate with absolute measures such as \ac{CTR}.
However, we note that in this study no more than six rankers were compared, thus such a study would likely miss a systematic error of 2.2\%.
In fact, \citet{chapelle2012large} note themselves that they cannot confidently claim team-draft interleaving is completely unbiased.
\citet{schuth2015predicting} performed a larger comparison involving 38 ranking systems, but again, too small to reliably detect a small systematic error.

It appears that the field is missing a large scale comparison that involves a large enough number of rankers to observe small systematic errors.
If such an error is found, the next step is to identify if certain types of ranking behavior are erroneously and systematically disfavored.
While these questions remain unanswered, we are concerned that the claims of unbiasedness in previous interleaving work (see Section~\ref{sec:related:interleaving}) give practitioners an unwarranted sense of reliability in interleaving.


\setlength{\tabcolsep}{0.05em}

{\renewcommand{\arraystretch}{0.2}
\begin{figure}[t]
\centering
\begin{tabular}{c l l}
&
 \multicolumn{1}{c}{ \hspace{0em} Yahoo Webscope}
&
 \multicolumn{1}{c}{ \hspace{0em} MSLR Web30k}
\\
\rotatebox[origin=lt]{90}{\hspace{0.5em} \small Team-Draft Interleaving}
&
\includegraphics[scale=0.315]{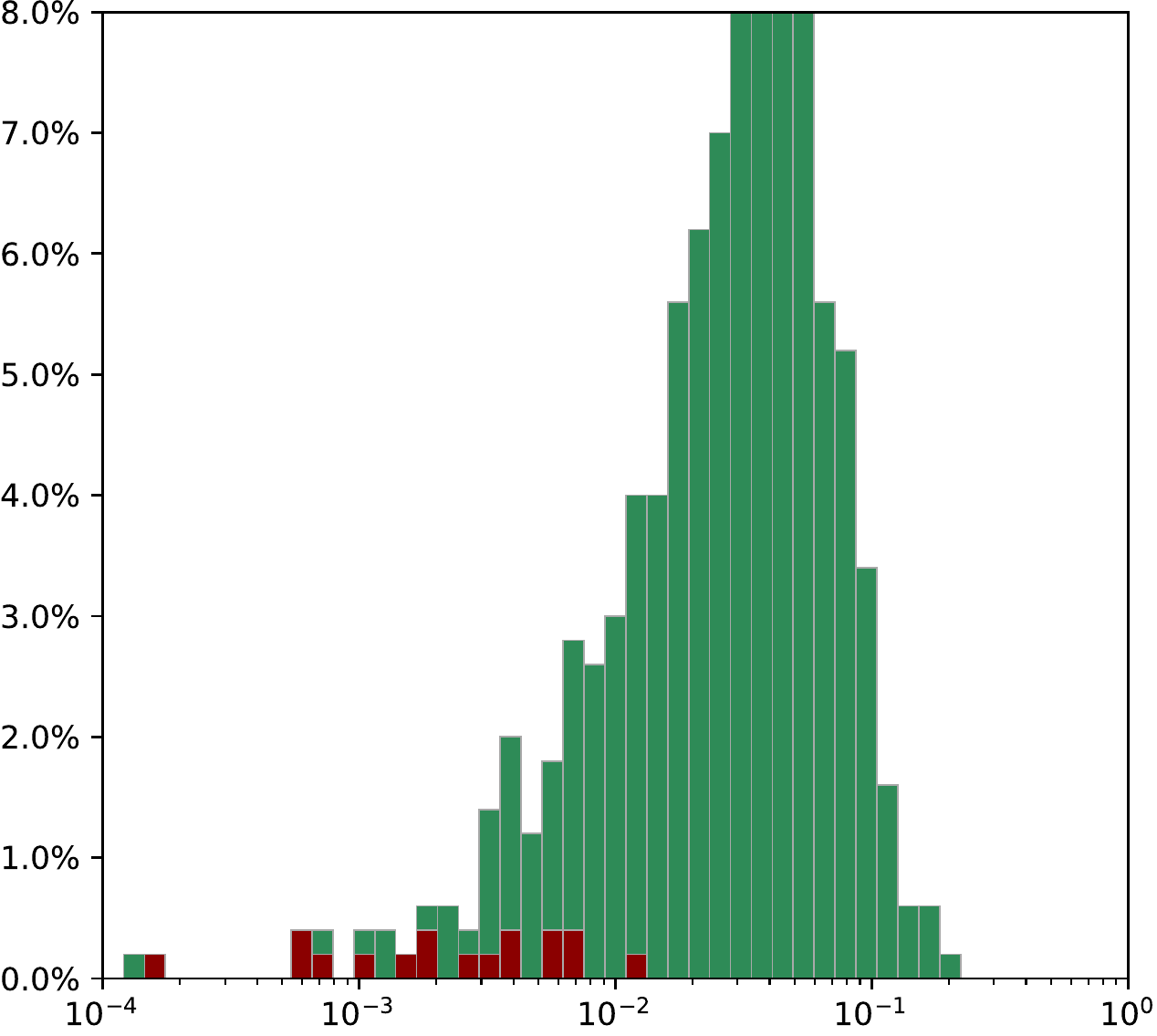} &
\includegraphics[scale=0.315]{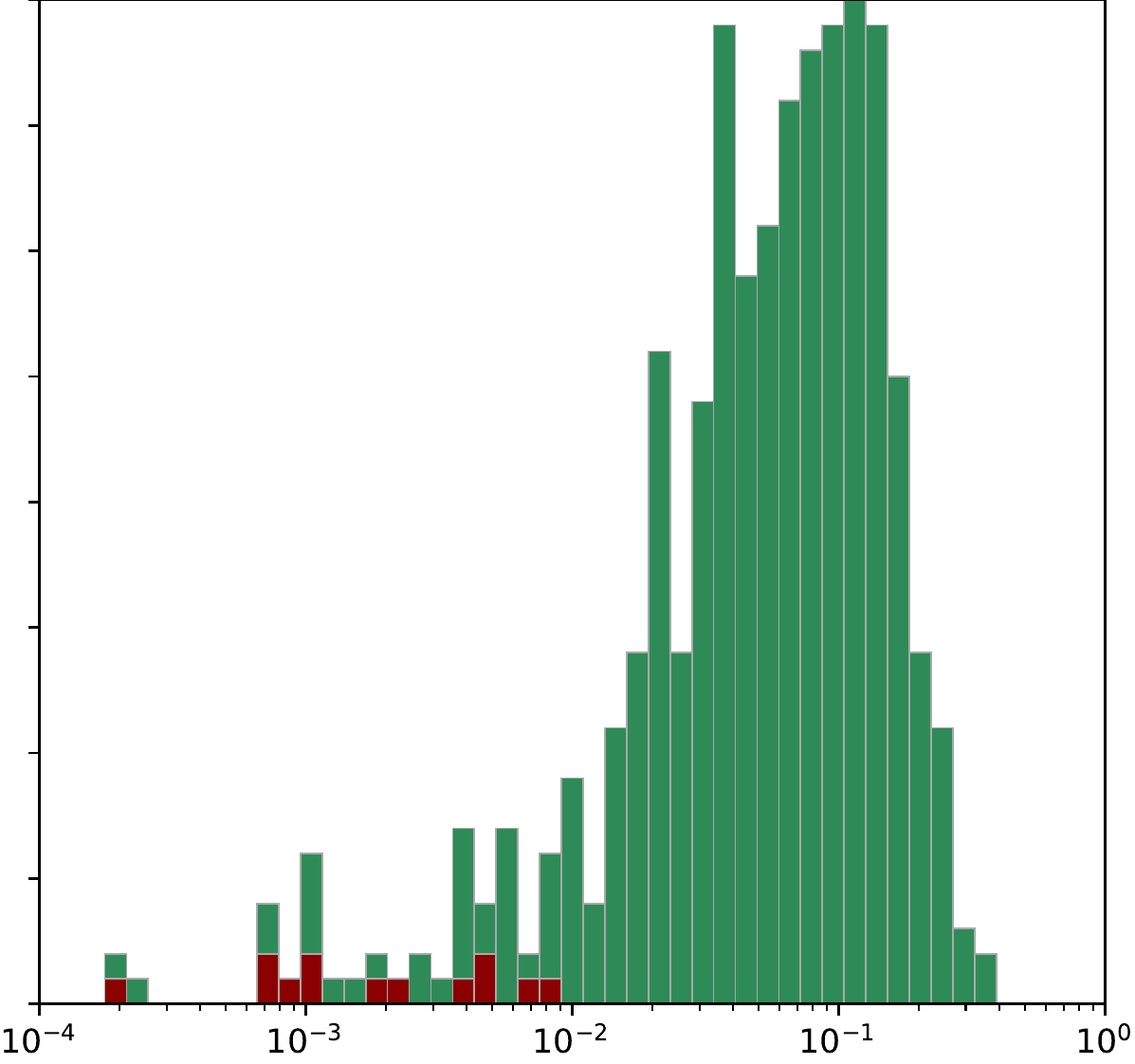} 
\\
\rotatebox[origin=lt]{90}{\hspace{0.1em} \small  Probabilistic Interleaving}
&
\includegraphics[scale=0.315]{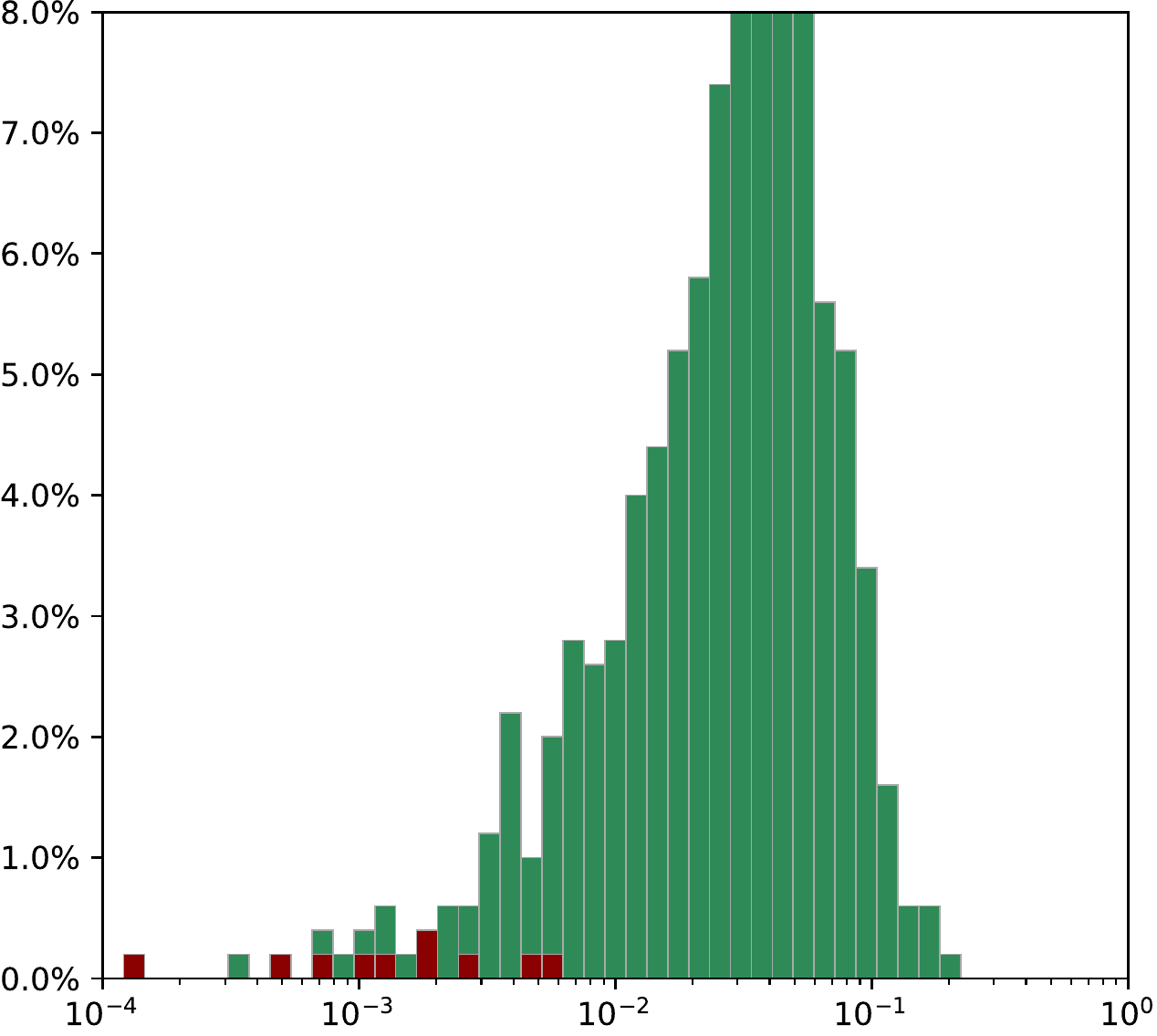} &
\includegraphics[scale=0.315]{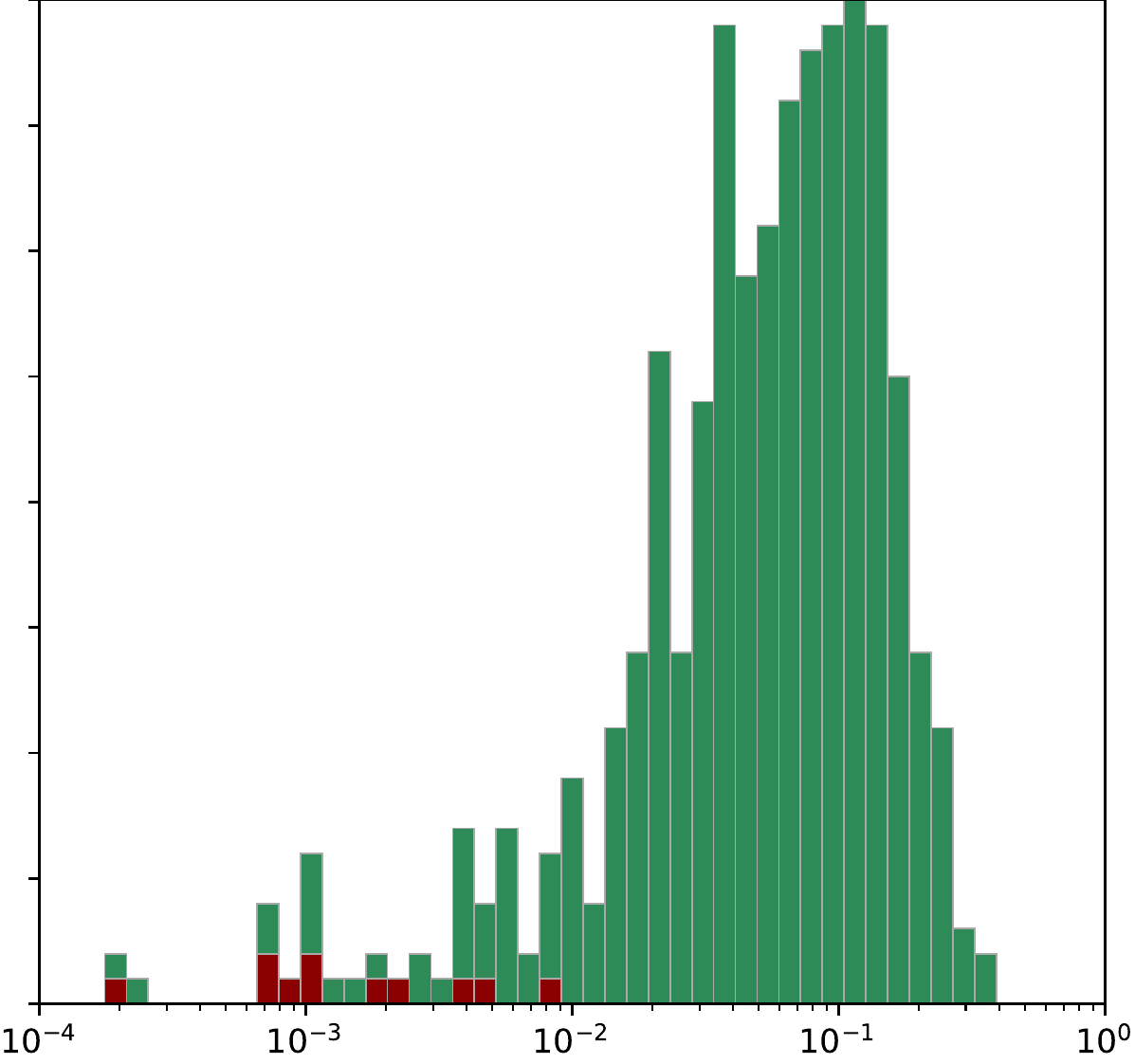} 
\\
\rotatebox[origin=lt]{90}{\hspace{0.5em} \small  Optimized Interleaving}
&
\includegraphics[scale=0.315]{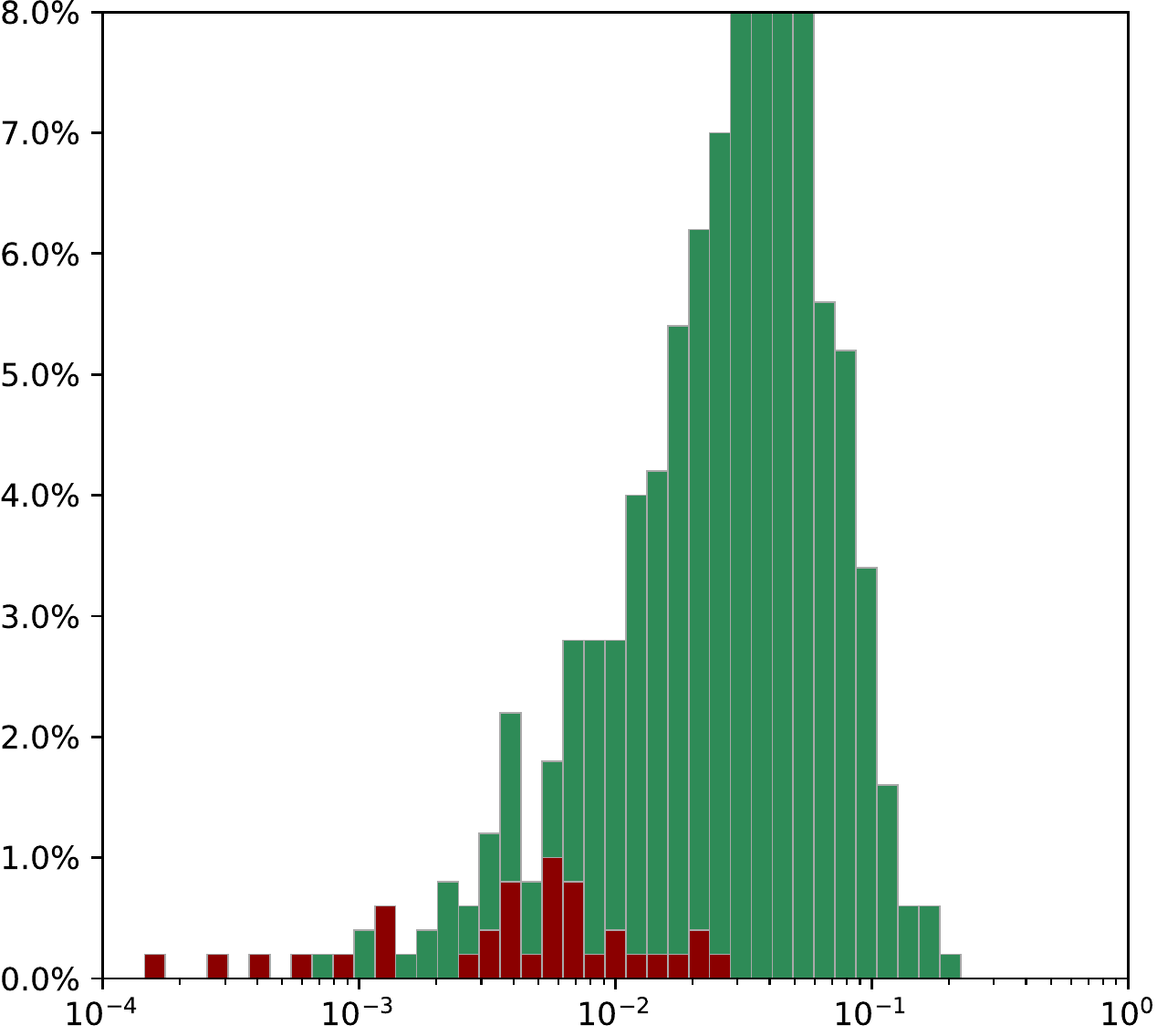} &
\includegraphics[scale=0.315]{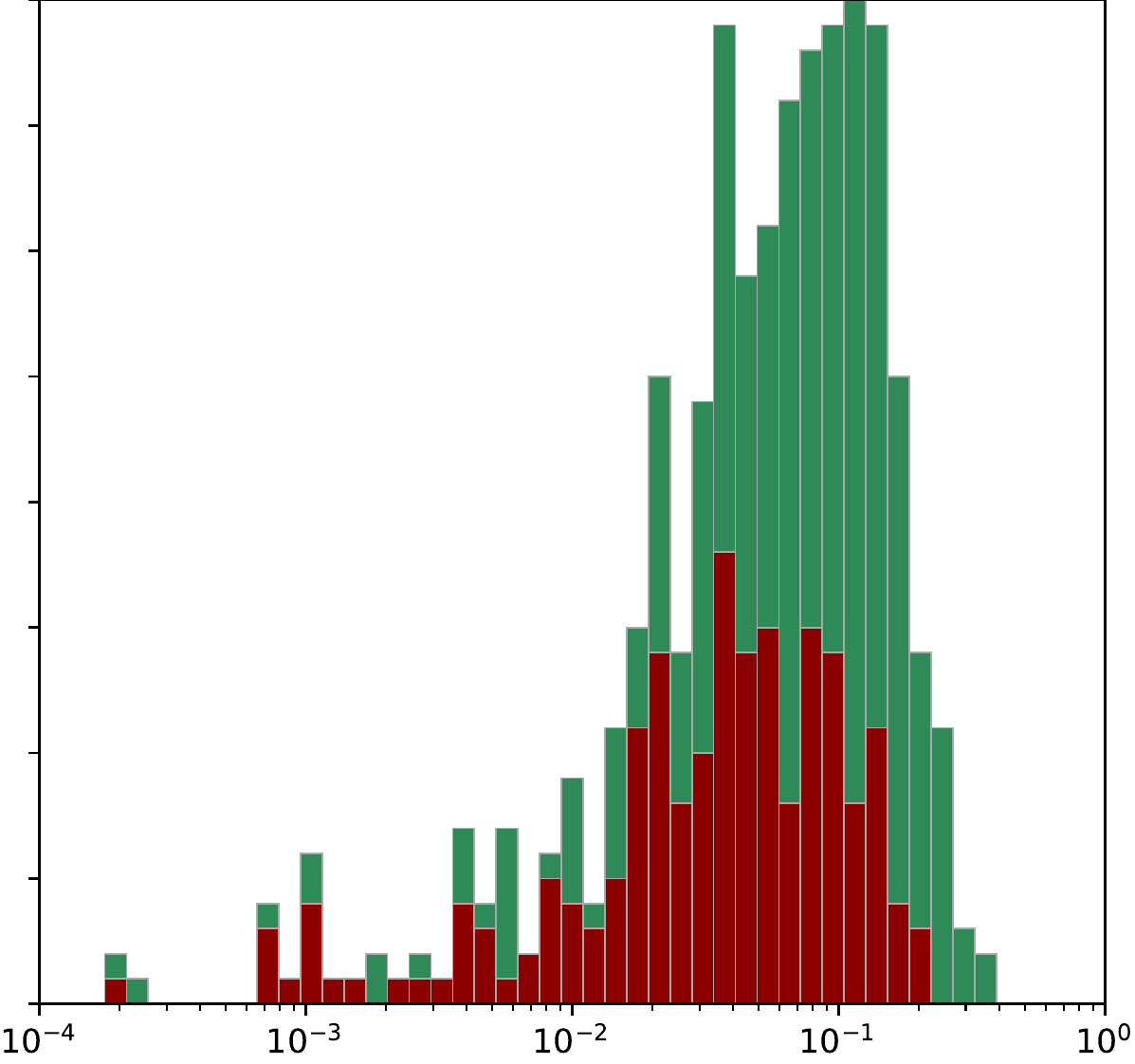} 
\\
\rotatebox[origin=lt]{90}{\hspace{3.5em} \small   A/B Testing}
\vspace{-0.5em}
&
\includegraphics[scale=0.315]{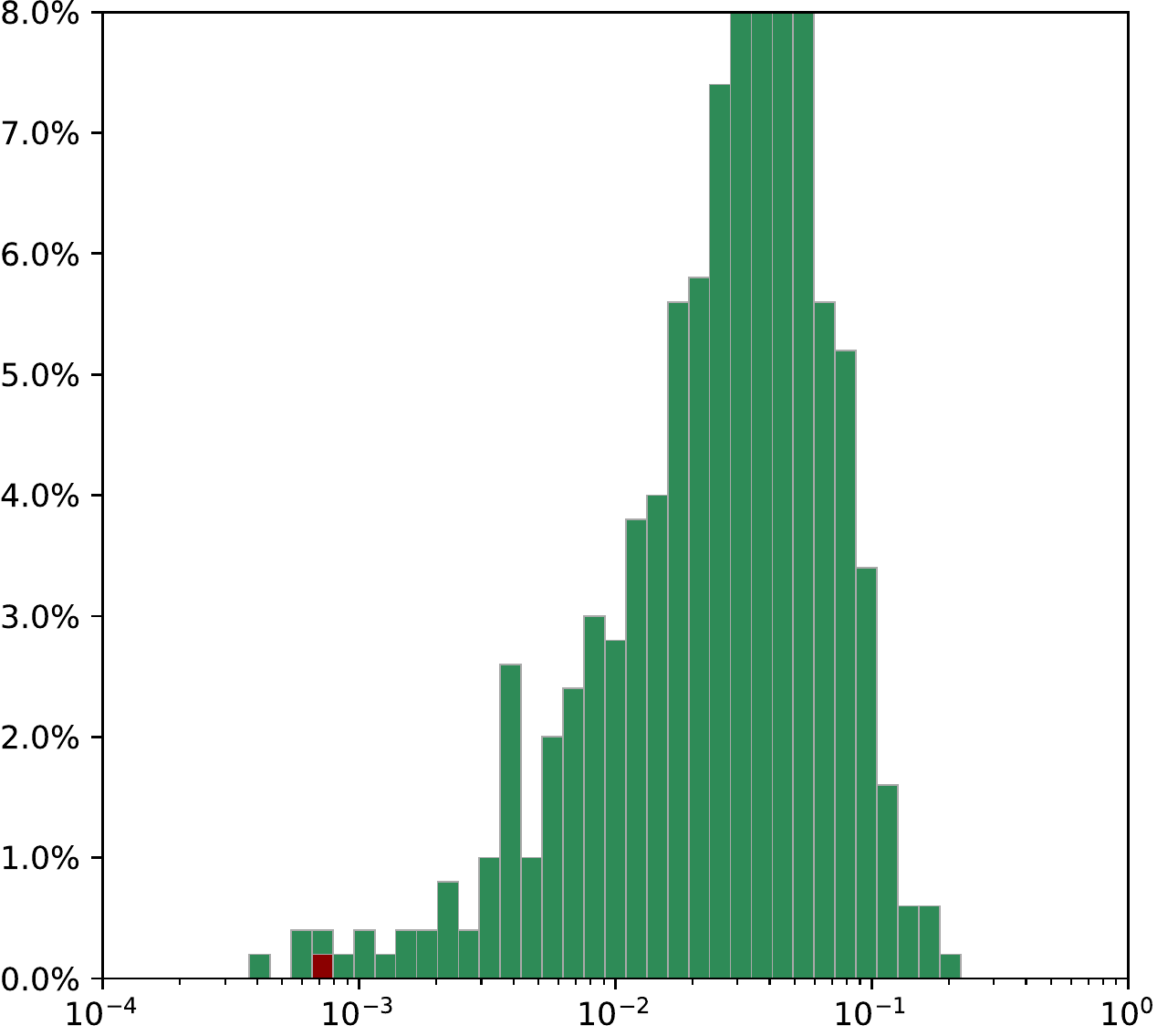} &
\includegraphics[scale=0.315]{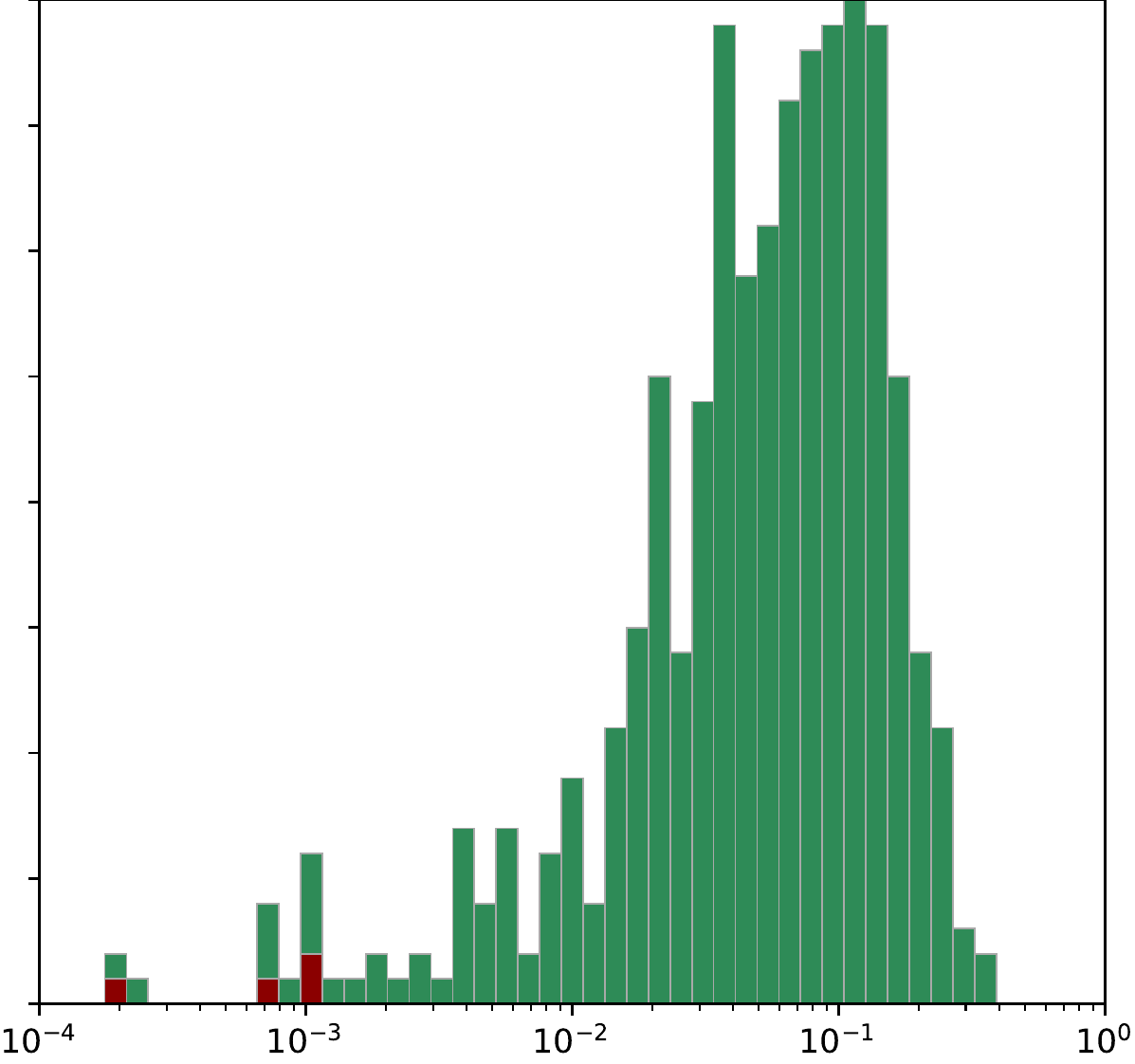} 
\\
\rotatebox[origin=lt]{90}{\hspace{0.3em} \small  LogOpt (Bias Estimated)}
&
\includegraphics[scale=0.315]{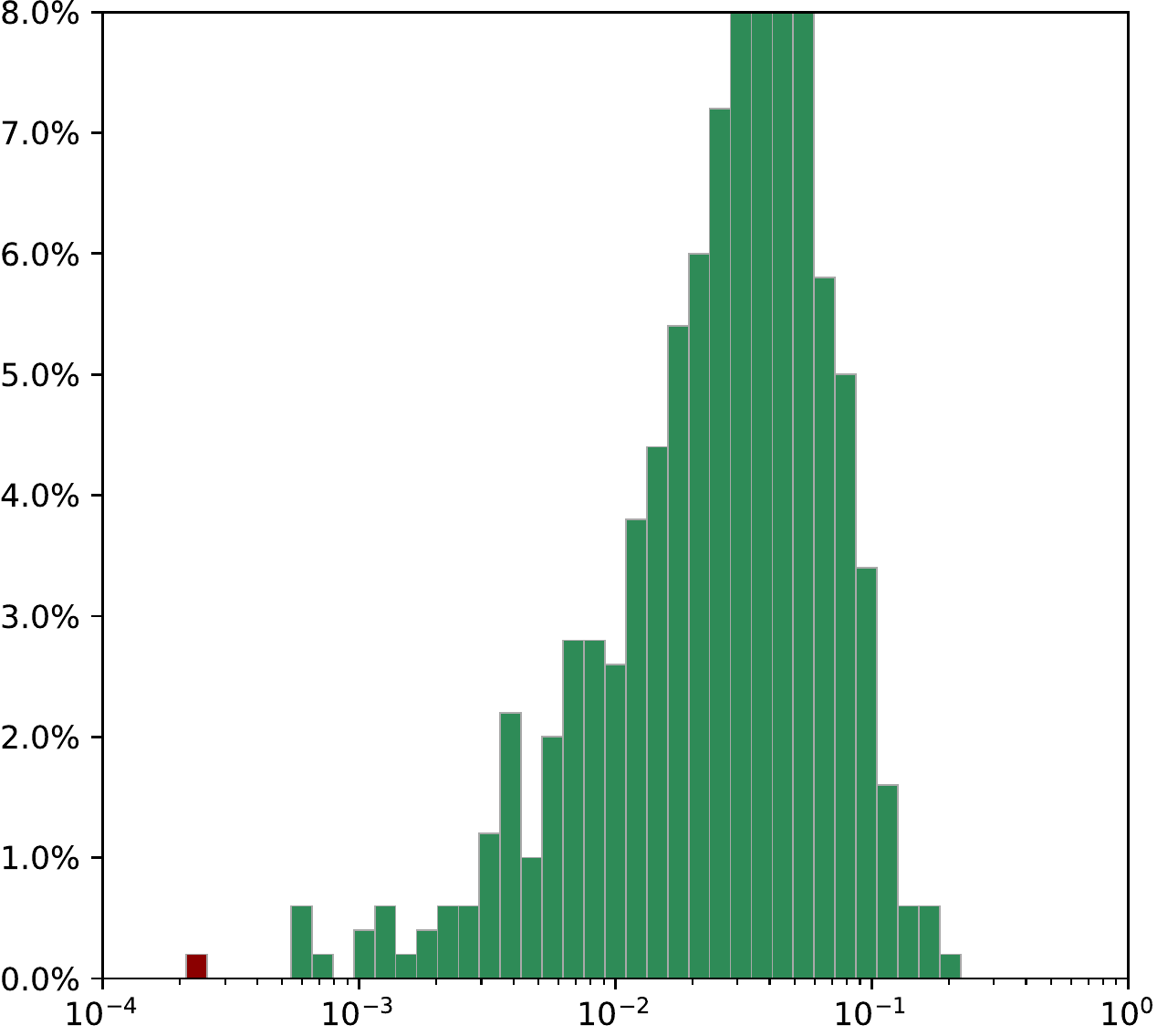} &
\includegraphics[scale=0.315]{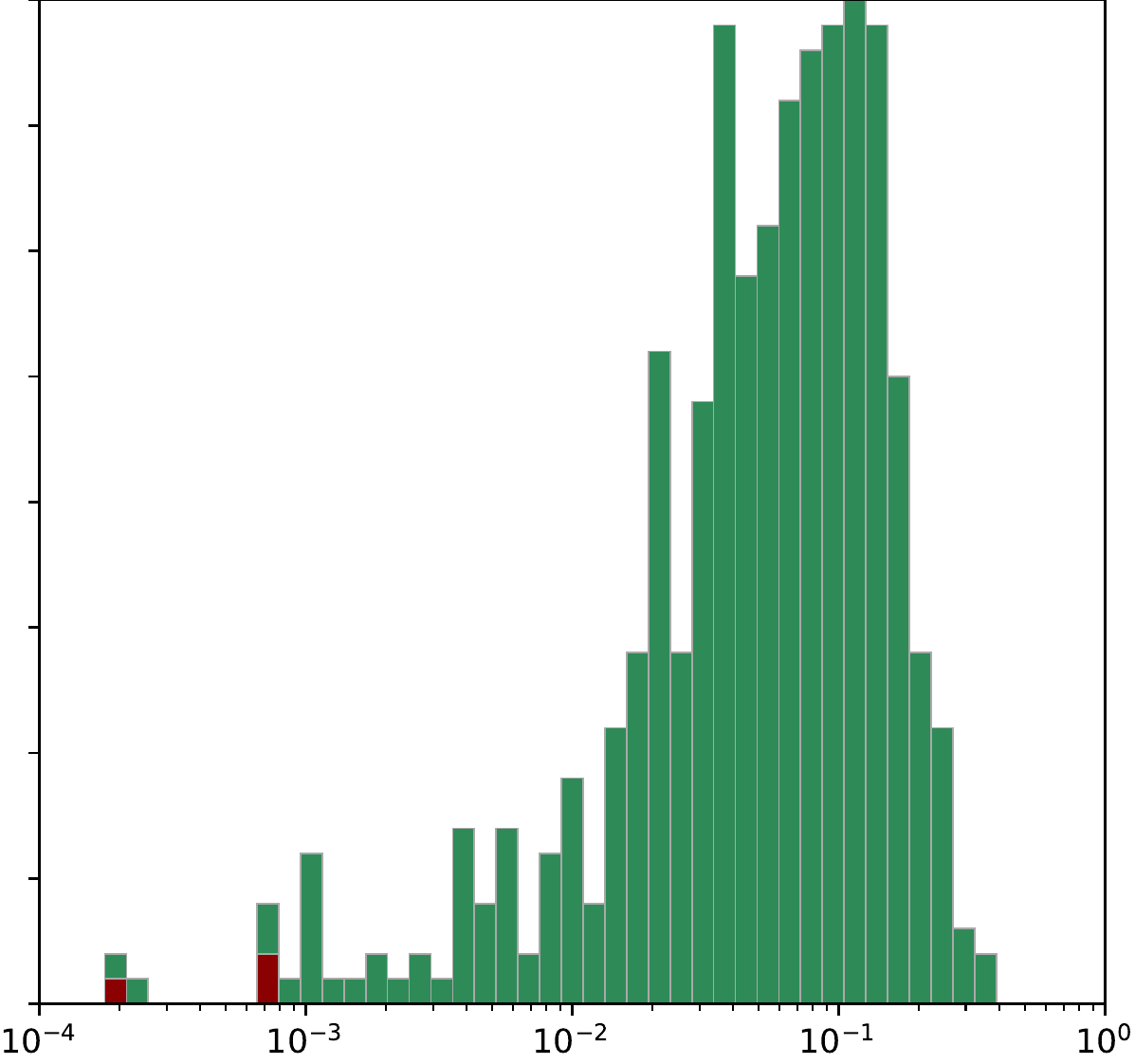} 
\\
&
 \multicolumn{1}{c}{\small  \hspace{1.8em} CTR difference}
&
 \multicolumn{1}{c}{\small  \hspace{1.8em} CTR difference}
\end{tabular}
\vspace{0.3\baselineskip}
\caption{
Distribution of errors over the \ac{CTR} differences of the rankers in the comparison; red indicates a binary error; green indicates a correctly inferred binary preference; results are on estimates based on $3 \cdot 10^6$ sampled queries.
}
\label{fig:errordistribution}
\end{figure}
}

%% file: sections/07-conclusion.tex

\section{Conclusion}

In this paper, we have introduced the \acf{LogOpt}: the first method that optimizes a logging policy for minimal variance counterfactual evaluation.
Counterfactual evaluation is proven to be unbiased w.r.t.\ position bias and item-selection bias under a wide range of logging policies.
With the introduction of \ac{LogOpt}, we now have an algorithm that can decide which rankings should be displayed for the fastest convergence.
Therefore, we argue that \ac{LogOpt} turns the existing counterfactual evaluation approach -- which is indifferent to the logging policy -- into an online approach -- which instructs the logging policy.

Our experimental results show that \ac{LogOpt} can lead to a better data-efficiency than A/B testing, without introducing the bias of interleaving.
While our findings are mostly theoretical, they do suggest that future work should further investigate the bias in interleaving methods.
Our results suggest that all interleaving methods make systematic errors, in particular when rankers with a similar \ac{CTR} are compared.
Furthermore, to the best of our knowledge, no empirical studies have been performed that could measure such a bias, our findings strongly show that such a study would be highly valuable to the field.
Finally, \ac{LogOpt} shows that in theory an evaluation method that is both unbiased and efficient is possible, if future work finds that these theoretical findings match empirical results with real users, this could be the start of a new line of theoretically-justified online evaluation methods.

%% file: sections/08-appendix.tex

\setlength{\tabcolsep}{1em}
\renewcommand{\arraystretch}{1.0}

\section{Proof of Bias in Interleaving}
\label{sec:appendix:bias}

Section~\ref{sec:related:interleaving} claimed that for the discussed interleaving methods, an example can be constructed so that in expectation the wrong binary outcome is estimated w.r.t.\ the actual expected \ac{CTR} differences.
These examples are enough to prove that these interleaving methods are biased w.r.t.\ \ac{CTR} differences.
In the following sections we will introduce a single example for each interleaving method.

For clarity, we will keep these examples as basic as possible.
We consider a ranking setting where only a single query $q_1$ occurs, i.e.\ $P(q_1) = 1$, furthermore, there are only three documents to be ranked: $A$, $B$, and $C$.
The two policies $\pi_1$ and $\pi_2$ in the comparison are both deterministic so that:
$\pi_1([A, B, C] \mid q_1) = 1$
and
$\pi_2([B, C, A] \mid q_1) = 1$.
Thus $\pi_1$ will always display the ranking: $[A, B, C]$, and $\pi_2$ the ranking: $[B, C, A]$.
Furthermore, document $B$ is completely non-relevant: $\zeta_B = 0$, consequently, $B$ can never receive clicks; this will make our examples even simpler.

The true $\mathbbm{E}[\text{CTR}]$ difference is thus:
\begin{equation}
\Delta(\pi_1, \pi_2) = (\theta_1 - \theta_3) \zeta_A + (\theta_3 - \theta_2) \zeta_C.
\end{equation}
For each interleaving method, will now show that position bias parameters $\theta_1$, $\theta_2$, and $\theta_3$ and relevances $\zeta_A$ and $\zeta_C$ exist where the wrong binary outcome is estimated.

\subsection{Team-Draft Interleaving}
Team-Draft Interleaving~\citep{radlinski2008does} lets rankers take turns to add their top document and keeps track which ranker added each document.
In total there are four possible interleaving and assignment combinations, each is equally probable:
\begin{table}[h]
\begin{tabular}{c c c c}
\toprule 
Interleaving & Ranking & Assignments & Probability \\
\midrule
$R_1$ & A, B, C & 1, 2, 1 & 1/4 \\
$R_2$ & A, B, C & 1, 2, 2 & 1/4 \\
$R_3$ & B, A, C & 2, 1, 1 & 1/4 \\
$R_4$ & B, A, C & 2, 1, 2 & 1/4 \\
\bottomrule
\end{tabular}
\end{table}

\noindent
Per issued query Team-Draft Interleaving produces a binary outcome, this is based on which ranker had most of its assigned documents clicked.
To match our \ac{CTR} estimate, we use $1$ to indicate $\pi_1$ receiving more clicks, and $-1$ for $\pi_2$.
Per interleaving we can compute the probability of each outcome:
\begin{align*}
P(\text{outcome} = \phantom{-}1 \mid R_1) &= \theta_1\zeta_A + (1- \theta_1\zeta_A)\theta_3\zeta_C,
\\
P(\text{outcome} = \phantom{-}1 \mid R_2) &= \theta_1\zeta_A(1- \theta_3\zeta_C),
\\
P(\text{outcome} = \phantom{-}1 \mid R_3) &= \theta_2\zeta_A + (1- \theta_2\zeta_A)\theta_3\zeta_C,
\\
P(\text{outcome} = \phantom{-}1 \mid R_4) &= \theta_2\zeta_A (1- \theta_3\zeta_C),
\\
P(\text{outcome} = -1 \mid R_1) &= 0,
\\
P(\text{outcome} = -1 \mid R_2) &= (1-\theta_1\zeta_A)\theta_3\zeta_C,
\\
P(\text{outcome} = -1 \mid R_3) &= 0,
\\
P(\text{outcome} = -1 \mid R_4) &= (1- \theta_2\zeta_A)\theta_3\zeta_C.
\end{align*}
Since every interleaving is equally likely, we can easily derive the unconditional probabilities:
\begin{align*}
P(\text{outcome} = \phantom{-}1) &= \frac{1}{4}\Big(
\theta_1\zeta_A + (1- \theta_1\zeta_A)\theta_3\zeta_C
+  \theta_1\zeta_A(1- \theta_3\zeta_C)
\\ & \quad \,\,
+ \theta_2\zeta_A + (1- \theta_2\zeta_A)\theta_3\zeta_C
+ \theta_2\zeta_A (1- \theta_3\zeta_C)
\Big),
\\
P(\text{outcome} = -1) &= \frac{1}{4}\Big(
(1-\theta_1\zeta_A)\theta_3\zeta_C
+ (1- \theta_2\zeta_A)\theta_3\zeta_C
\Big).
\end{align*}
With these probabilities, the expected outcome is straightforward to calculate:
\begin{align*}
\mathbbm{E}[\text{outcome}] &=
 P(\text{outcome} = 1) - P(\text{outcome} = -1)
 \\ &=
 \frac{1}{4}\Big(
\theta_1\zeta_A
+  \theta_1\zeta_A(1- \theta_3\zeta_C)
+ \theta_2\zeta_A 
+ \theta_2\zeta_A (1- \theta_3\zeta_C)
\Big)
\\
&> 0.
\end{align*}
Interestingly, without knowing the values for $\theta$, $\zeta_A$ and $\zeta_C$, we already know that the expected outcome is positive.
Therefore, we can simply choose values that lead to a negative \ac{CTR} difference, and the expected outcome will be incorrect.
For this example, we choose the position bias:
$\theta_1 = 1.0$,
$\theta_2 = 0.9$, and
$\theta_3 = 0.8$;
and the relevances:
$\zeta_1 = 0.1$, and
$\zeta_3 = 1.0$.
As a result, the expected binary outcome of Team-Draft Interleaving will not match the true $\mathbbm{E}[\text{CTR}]$ difference:
\begin{equation}
\Delta(\pi_1, \pi_2) < 0 \land \mathbbm{E}[\text{outcome}]  > 0.
\end{equation}
Therefore, we have proven that Team-Draft Interleaving is biased w.r.t.\ \ac{CTR} differences.

\subsection{Probabilistic Interleaving}
Probabilistic Interleaving~\citep{hofmann2011probabilistic} treats rankings as distributions over documents, we follow the soft-max approach of~\citet{hofmann2011probabilistic} and use $\tau=4.0$ as suggested.
Probabilistic Interleaving creates interleavings by sampling randomly from one of the rankings, unlike Team-Draft Interleaving it does not remember which ranking added each document.
Because rankings are treated as distributions, every possible permutation is a valid interleaving, leading to six possibilities with different probabilities of being displayed.
When clicks are received, every possible assignment is considered and the expected outcome is computed over all possible assignments.
Because there are 36 possible rankings and assignment combinations, we only report every possible ranking and the probabilities for documents $A$ or $C$ being added by $\pi_1$:
\setlength{\tabcolsep}{0.4em}
\begin{table}[h]
\begin{tabular}{c c c c c}
\toprule 
Interl. & Ranking & $P(\text{add}(A) = 1)$ & $P(\text{add}(C) = 1)$ & Probability \\
\midrule
$R_1$ & A, B, C & 0.9878 & 0.4701 & 0.4182 \\
$R_2$ & A, C, B & 0.9878 & 0.4999 & 0.0527 \\
$R_3$ & B, A, C & 0.8569 & 0.0588 & 0.2849 \\
$R_4$ & B, C, A & 0.5000 & 0.0588 & 0.2094 \\
$R_5$ & C, A, B & 0.9872 & 0.5000 & 0.0166 \\
$R_6$ & C, B, A & 0.5000 & 0.0562 & 0.0182 \\
\bottomrule
\end{tabular}
\end{table}

\noindent
These probabilities are enough to compute the expected outcome, similar as the procedure we used for Team-Draft Interleaving.
We will not display the full calculation here as it is extremely long; we recommend using some form of computer assistance to perform these calculations.
While there are many possibilities, we choose the following position bias:
$\theta_1 = 1.0$,
$\theta_2 = 0.9$, and
$\theta_3 = 0.3$;
and relevance:
$\zeta_1 = 0.5$, and
$\zeta_3 = 1.0$.
This leads to the following erroneous result:
\begin{equation}
\Delta(\pi_1, \pi_2) < 0 \land \mathbbm{E}[\text{outcome}]  > 0.
\end{equation}
Therefore, we have proven that Probabilistic Interleaving is biased w.r.t.\ \ac{CTR} differences.

\subsection{Optimized Interleaving}
Optimized Interleaving casts interleaving as an optimization problem~\citep{radlinski2013optimized}.
Optimized Interleaving works with a credit function: each clicked document produces a positive or negative credit.
The sum of all credits is the final estimated outcome.
We follow \citet{radlinski2013optimized} and use the linear rank difference, resulting in the following credits per document:
$\text{click-credit}(A) = 2$,
$\text{click-credit}(B) = -1$, and
$\text{click-credit}(C) = -1$.
Then the set of allowed interleavings is created, these are all the rankings that do not contradict a pairwise document preference that both rankers agree on.
Given this set of interleavings, a distribution over them is found so that if every document is equally relevant then no preference is found.\footnote{\citet{radlinski2013optimized} state that if clicks are not correlated with relevance then no preference should be found, in their click model (and ours) these two requirements are actually equivalent.}
For our example, the only valid distribution over interleavings is the following:
\setlength{\tabcolsep}{0.4em}
\begin{table}[h]
\begin{tabular}{c c c}
\toprule 
Interleaving & Ranking & Probability \\
\midrule
$R_1$ & A, B, C & $1/3$ \\
$R_2$ & B, A, C & $1/3$ \\
$R_3$ & B, C, A & $1/3$ \\
\bottomrule
\end{tabular}
\end{table}

\noindent
The expected credit outcome shows us which ranker will be preferred in expectation:
\begin{equation}
\mathbbm{E}[\text{credit}] = \frac{1}{3}\big( 2(\theta_1 + \theta_2 + \theta_3)\zeta_A - (\theta_2 + 2\theta_3)\zeta_C \big).
\end{equation}
We choose the position bias:
$\theta_1 = 1.0$,
$\theta_2 = 0.9$, and
$\theta_3 = 0.9$;
and the relevances:
$\zeta_1 = 0.5$,
$\zeta_3 = 1.0$.
As a result, the true $\mathbbm{E}[\text{CTR}]$ difference is positive, but optimized interleaving will prefer $\pi_2$ in expectation:
\begin{equation}
\Delta(\pi_1, \pi_2) > 0 \land \mathbbm{E}[\text{credit}] < 0.
\end{equation}
Therefore, we have proven that Optimized Interleaving is biased w.r.t.\ \ac{CTR} differences.

\section{Expanded Explanation of Gradient Approximation}
\label{sec:appendix:approx}

This section describes our Monte-Carlo approximation of the variance gradient in more detail.
We repeat the steps described in Section~\ref{sec:method:derivates} as well as some additional intermediate steps, this should make it easier for a reader to verify our theory.

First, we assume that policies place the documents in order of rank and the probability of placing an individual document at rank $x$ only depends on the previously placed documents.
Let $R_{1:x-1}$ indicate the (incomplete) ranking from rank $1$ up to rank $x$, then $\pi_0(d \mid R_{1:x-1}, q)$ indicates the probability that document $d$ is placed at rank $x$ given that the ranking up to $x$ is $R_{1:x-1}$.
The probability of a ranking $R$ of length $K$ is thus:
\begin{equation}
\pi_0(R \mid q) = \prod_{x=1}^{K} \pi_0(R_x \mid R_{1:x-1}, q).
\end{equation}
The probability of a ranking $R$ up to rank $k$ is:
\begin{equation}
\pi_0(R_{1:k} \mid q) = \prod_{x=1}^{k} \pi_0(R_x \mid R_{1:x-1}, q).
\end{equation}
Therefore the propensity (cf.\ Eq.~\ref{eq:prop}) can be rewritten to:
\begin{equation}
\rho(d \,|\, q) = \sum_{k=1}^K \theta_k \sum_R \pi_0(R_{1:k-1} \mid q) \pi_0(d \,|\, R_{1:k-1}, q).
\end{equation}
Before we take the gradient of the propensity, we note that the gradient of the probability of a single ranking is:
\begin{equation}
\frac{\delta \pi_0(R \mid q) }{\delta \pi_0}
 =
\sum_{x=1}^K
  \frac{\pi_0(R \mid q)}{\pi_0(R_x \mid R_{1:x},  q)}
  \left[\frac{\delta \pi_0(R_x \mid R_{1:x-1},  q) }{\delta \pi_0} \right].
  \label{eq:appendix:rankingprob}
\end{equation}
Using this gradient, we can derive the gradient of the propensity w.r.t.\ the policy: 
\begin{align}
\frac{\delta \rho(d |\, q)}{\delta \pi_0} &=  \sum_{k=1}^K \theta_k \sum_R \pi_0(R_{1:k-1} |\,  q) 
\Bigg(\left[ \frac{\delta \pi_0(d |\, R_{1:k-1}, q)}{\delta \pi_0} \right] \nonumber
\\
& \quad   + 
\sum_{x=1}^{k-1} \frac{\pi_0(d |\, R_{1:k-1}, q)}{\pi_0(R_x |\, R_{1:x-1} , q)}\left[ \frac{\delta\pi_0(R_x |\,  R_{1:x-1} , q)}{\delta \pi_0} \right]
\Bigg).
\end{align}
To avoid iterating over all rankings in the $\sum_R$ sum, we sample $M$ rankings:
$R^m \sim \pi_0(R \mid q)$, and a click pattern on each ranking: $c^m \sim P(c \mid R^m)$.
This enables us to make the following approximation:
\begin{align}
\widehat{\rho\text{-grad}}(d)
&=
 \frac{1}{M} \sum_{m=1}^{M} \sum_{k=1}^K \theta_k 
\Bigg(
\left[ \frac{\delta \pi_0(d |\, R^m_{1:k-1}, q)}{\delta \pi_0} \right] 
\\
& \quad \quad \, + 
\sum_{x=1}^{k-1} \frac{\pi_0(d |\, R^m_{1:k-1}, q)}{\pi_0(R^m_x  |\, R^m_{1:x-1} , q)}\left[ \frac{\delta\pi_0(R^m_x  |\, R^m_{1:x-1} , q)}{\delta \pi_0} \right]
\Bigg),
\nonumber
\end{align}
since $\frac{\delta \rho(d |\, q)}{\delta \pi_0} \approx \widehat{\rho\text{-grad}}(d, q)$.
The second part of Eq.~\ref{eq:gradient} is:
\begin{equation}
\Bigg[ \frac{\delta}{\delta \pi_0} \bigg(\Delta - \sum_{d : c(d) = 1}\frac{\lambda_d}{\rho_d}\bigg)^2 \Bigg]
= 2\bigg(\Delta - \sum_{d : c(d) = 1}\frac{\lambda_d}{\rho_d}\bigg) \sum_{d : c(d) = 1}\frac{\lambda_d}{\rho_d^2} \left[ \frac{\delta \rho_d}{\delta \pi_0} \right],
\end{equation}
using $\widehat{\rho\text{-grad}}(d)$ we get the approximation: 
\begin{equation}
\widehat{\text{error-grad}}(c) =
2\bigg(\Delta - \sum_{d : c(d) = 1}\frac{\lambda_d}{\rho_d}\bigg) \sum_{d : c(d) = 1}\frac{\lambda_d}{\rho_d^2} \widehat{\rho\text{-grad}}(d).
\end{equation}
Next, we consider the gradient of a single click pattern: 
\begin{equation}
\frac{\delta}{\delta \pi_0} P(c \mid q) = \sum_R P(c \mid R) \left[\frac{\delta \pi_0(R \mid q)}{\delta \pi_0} \right].
\end{equation}
This can then be used to reformulate the first part of Eq.~\ref{eq:gradient}:
\begin{align}
\sum_c&
\left[\frac{\delta}{\delta \pi_0}P(c \mid q)\right] \bigg(\Delta - \sum_{d : c(d) = 1}\frac{\lambda_d}{\rho_d}\bigg)^2
\nonumber \\ 
&= \sum_c
\sum_R P(c \mid R) \left[\frac{\delta \pi_0(R \mid q)}{\delta \pi_0}\right] \bigg(\Delta - \sum_{d : c(d) = 1}\frac{\lambda_d}{\rho_d}\bigg)^2
\end{align}
Making use of Eq.~\ref{eq:appendix:rankingprob}, we approximate this with:
\begin{align}
&\widehat{\text{freq-grad}}(R, c) = \\
& \bigg(\Delta - \sum_{d : c(d) = 1}\frac{\lambda_d}{\rho_d}\bigg)^2
 \sum_{x=1}^K
\frac{1}{\pi_0(R_x \mid R_{1:x-1},  q)}
\left[\frac{\delta \pi_0(R_x \mid R_{1:x-1},  q) }{\delta \pi_0} \right].
\nonumber
\end{align}
Combining the approximation of both parts of Eq.~\ref{eq:gradient}, allows us to approximate the complete gradient:
\begin{equation}
\begin{split}
&\frac{\delta \text{Var}(\hat{\Delta}_{IPS}^{\pi_0} \mid q)}{\delta \pi_0} 
\approx{}\\
&\mbox{}\quad
\frac{1}{M}\sum_{m=1}^M \widehat{\text{freq-grad}}(R^m, c^m) + \widehat{\text{error-grad}}(c^m).
\end{split}
\end{equation}
This completes our expanded description of the gradient approximation.
We have shown that we can approximate the gradient of the variance w.r.t.\ a logging policy $\pi_0$, based on rankings sampled from $\pi_0$ and our current estimated click model $\hat{\theta}$, $\hat{\zeta}$, while staying computationally feasible.

%% file: 2020-online-counterfactual-evaluation.bbl

\begin{thebibliography}{24}


\ifx \showCODEN    \undefined \def \showCODEN     #1{\unskip}     \fi
\ifx \showDOI      \undefined \def \showDOI       #1{#1}\fi
\ifx \showISBNx    \undefined \def \showISBNx     #1{\unskip}     \fi
\ifx \showISBNxiii \undefined \def \showISBNxiii  #1{\unskip}     \fi
\ifx \showISSN     \undefined \def \showISSN      #1{\unskip}     \fi
\ifx \showLCCN     \undefined \def \showLCCN      #1{\unskip}     \fi
\ifx \shownote     \undefined \def \shownote      #1{#1}          \fi
\ifx \showarticletitle \undefined \def \showarticletitle #1{#1}   \fi
\ifx \showURL      \undefined \def \showURL       {\relax}        \fi
\providecommand\bibfield[2]{#2}
\providecommand\bibinfo[2]{#2}
\providecommand\natexlab[1]{#1}
\providecommand\showeprint[2][]{arXiv:#2}

\bibitem[\protect\citeauthoryear{Agarwal, Zaitsev, Wang, Li, Najork, and
  Joachims}{Agarwal et~al\mbox{.}}{2019}]%
        {agarwal2019estimating}
\bibfield{author}{\bibinfo{person}{Aman Agarwal}, \bibinfo{person}{Ivan
  Zaitsev}, \bibinfo{person}{Xuanhui Wang}, \bibinfo{person}{Cheng Li},
  \bibinfo{person}{Marc Najork}, {and} \bibinfo{person}{Thorsten Joachims}.}
  \bibinfo{year}{2019}\natexlab{}.
\newblock \showarticletitle{Estimating Position Bias without Intrusive
  Interventions}. In \bibinfo{booktitle}{\emph{WSDM}}. ACM,
  \bibinfo{pages}{474--482}.
\newblock


\bibitem[\protect\citeauthoryear{Ai, Bi, Luo, Guo, and Croft}{Ai
  et~al\mbox{.}}{2018}]%
        {ai2018unbiased}
\bibfield{author}{\bibinfo{person}{Qingyao Ai}, \bibinfo{person}{Keping Bi},
  \bibinfo{person}{Cheng Luo}, \bibinfo{person}{Jiafeng Guo}, {and}
  \bibinfo{person}{W~Bruce Croft}.} \bibinfo{year}{2018}\natexlab{}.
\newblock \showarticletitle{Unbiased Learning to Rank with Unbiased Propensity
  Estimation}. In \bibinfo{booktitle}{\emph{SIGIR}}. ACM,
  \bibinfo{pages}{385--394}.
\newblock


\bibitem[\protect\citeauthoryear{Chapelle and Chang}{Chapelle and
  Chang}{2011}]%
        {Chapelle2011}
\bibfield{author}{\bibinfo{person}{Olivier Chapelle} {and} \bibinfo{person}{Yi
  Chang}.} \bibinfo{year}{2011}\natexlab{}.
\newblock \showarticletitle{{Yahoo! Learning to Rank Challenge Overview}}.
\newblock \bibinfo{journal}{\emph{Journal of Machine Learning Research}}
  \bibinfo{volume}{14} (\bibinfo{year}{2011}), \bibinfo{pages}{1--24}.
\newblock


\bibitem[\protect\citeauthoryear{Chapelle, Joachims, Radlinski, and
  Yue}{Chapelle et~al\mbox{.}}{2012}]%
        {chapelle2012large}
\bibfield{author}{\bibinfo{person}{Olivier Chapelle}, \bibinfo{person}{Thorsten
  Joachims}, \bibinfo{person}{Filip Radlinski}, {and} \bibinfo{person}{Yisong
  Yue}.} \bibinfo{year}{2012}\natexlab{}.
\newblock \showarticletitle{Large-scale Validation and Analysis of Interleaved
  Search Evaluation}.
\newblock \bibinfo{journal}{\emph{ACM Transactions on Information Systems
  (TOIS)}} \bibinfo{volume}{30}, \bibinfo{number}{1} (\bibinfo{year}{2012}),
  \bibinfo{pages}{1--41}.
\newblock


\bibitem[\protect\citeauthoryear{Chuklin, Markov, and de~Rijke}{Chuklin
  et~al\mbox{.}}{2015}]%
        {chuklin2015click}
\bibfield{author}{\bibinfo{person}{Aleksandr Chuklin}, \bibinfo{person}{Ilya
  Markov}, {and} \bibinfo{person}{Maarten de Rijke}.}
  \bibinfo{year}{2015}\natexlab{}.
\newblock \bibinfo{booktitle}{\emph{Click Models for Web Search}}.
\newblock \bibinfo{publisher}{Morgan \& Claypool Publishers}.
\newblock


\bibitem[\protect\citeauthoryear{Craswell, Zoeter, Taylor, and Ramsey}{Craswell
  et~al\mbox{.}}{2008}]%
        {craswell2008experimental}
\bibfield{author}{\bibinfo{person}{Nick Craswell}, \bibinfo{person}{Onno
  Zoeter}, \bibinfo{person}{Michael Taylor}, {and} \bibinfo{person}{Bill
  Ramsey}.} \bibinfo{year}{2008}\natexlab{}.
\newblock \showarticletitle{An Experimental Comparison of Click Position-bias
  Models}. In \bibinfo{booktitle}{\emph{WSDM}}. \bibinfo{pages}{87--94}.
\newblock


\bibitem[\protect\citeauthoryear{Fang, Agarwal, and Joachims}{Fang
  et~al\mbox{.}}{2019}]%
        {fang2019intervention}
\bibfield{author}{\bibinfo{person}{Zhichong Fang}, \bibinfo{person}{Aman
  Agarwal}, {and} \bibinfo{person}{Thorsten Joachims}.}
  \bibinfo{year}{2019}\natexlab{}.
\newblock \showarticletitle{Intervention Harvesting for Context-dependent
  Examination-bias Estimation}. In \bibinfo{booktitle}{\emph{SIGIR}}.
  \bibinfo{pages}{825--834}.
\newblock


\bibitem[\protect\citeauthoryear{Hofmann, Li, and Radlinski}{Hofmann
  et~al\mbox{.}}{2016}]%
        {hofmann2016online}
\bibfield{author}{\bibinfo{person}{Katja Hofmann}, \bibinfo{person}{Lihong Li},
  {and} \bibinfo{person}{Filip Radlinski}.} \bibinfo{year}{2016}\natexlab{}.
\newblock \showarticletitle{Online Evaluation for Information Retrieval}.
\newblock \bibinfo{journal}{\emph{Foundations and Trends in Information
  Retrieval}} \bibinfo{volume}{10}, \bibinfo{number}{1} (\bibinfo{year}{2016}),
  \bibinfo{pages}{1--117}.
\newblock


\bibitem[\protect\citeauthoryear{Hofmann, Whiteson, and de~Rijke}{Hofmann
  et~al\mbox{.}}{2011}]%
        {hofmann2011probabilistic}
\bibfield{author}{\bibinfo{person}{Katja Hofmann}, \bibinfo{person}{Shimon
  Whiteson}, {and} \bibinfo{person}{Maarten de Rijke}.}
  \bibinfo{year}{2011}\natexlab{}.
\newblock \showarticletitle{A Probabilistic Method for Inferring Preferences
  from Clicks}. In \bibinfo{booktitle}{\emph{CIKM}}. ACM,
  \bibinfo{pages}{249--258}.
\newblock


\bibitem[\protect\citeauthoryear{Hofmann, Whiteson, and Rijke}{Hofmann
  et~al\mbox{.}}{2013}]%
        {hofmann2013fidelity}
\bibfield{author}{\bibinfo{person}{Katja Hofmann}, \bibinfo{person}{Shimon
  Whiteson}, {and} \bibinfo{person}{Maarten~De Rijke}.}
  \bibinfo{year}{2013}\natexlab{}.
\newblock \showarticletitle{Fidelity, Soundness, and Efficiency of Interleaved
  Comparison Methods}.
\newblock \bibinfo{journal}{\emph{ACM Transactions on Information Systems
  (TOIS)}} \bibinfo{volume}{31}, \bibinfo{number}{4} (\bibinfo{year}{2013}),
  \bibinfo{pages}{1--43}.
\newblock


\bibitem[\protect\citeauthoryear{Joachims}{Joachims}{2003}]%
        {joachims2003evaluating}
\bibfield{author}{\bibinfo{person}{Thorsten Joachims}.}
  \bibinfo{year}{2003}\natexlab{}.
\newblock \showarticletitle{Evaluating Retrieval Performance Using Clickthrough
  Data}.
\newblock In \bibinfo{booktitle}{\emph{Text Mining}},
  \bibfield{editor}{\bibinfo{person}{J.~Franke},
  \bibinfo{person}{G.~Nakhaeizadeh}, {and} \bibinfo{person}{I.~Renz}} (Eds.).
  \bibinfo{publisher}{Physica Verlag}.
\newblock


\bibitem[\protect\citeauthoryear{Joachims, Granka, Pan, Hembrooke, and
  Gay}{Joachims et~al\mbox{.}}{2005}]%
        {joachims2017accurately}
\bibfield{author}{\bibinfo{person}{Thorsten Joachims}, \bibinfo{person}{Laura
  Granka}, \bibinfo{person}{Bing Pan}, \bibinfo{person}{Helene Hembrooke},
  {and} \bibinfo{person}{Geri Gay}.} \bibinfo{year}{2005}\natexlab{}.
\newblock \showarticletitle{Accurately Interpreting Clickthrough Data as
  Implicit Feedback}. In \bibinfo{booktitle}{\emph{SIGIR}}.
  \bibinfo{publisher}{ACM}, \bibinfo{pages}{154--161}.
\newblock


\bibitem[\protect\citeauthoryear{Joachims, Swaminathan, and Schnabel}{Joachims
  et~al\mbox{.}}{2017}]%
        {joachims2017unbiased}
\bibfield{author}{\bibinfo{person}{Thorsten Joachims}, \bibinfo{person}{Adith
  Swaminathan}, {and} \bibinfo{person}{Tobias Schnabel}.}
  \bibinfo{year}{2017}\natexlab{}.
\newblock \showarticletitle{Unbiased Learning-to-Rank with Biased Feedback}. In
  \bibinfo{booktitle}{\emph{WSDM}}. ACM, \bibinfo{pages}{781--789}.
\newblock


\bibitem[\protect\citeauthoryear{Kohavi and Longbotham}{Kohavi and
  Longbotham}{2017}]%
        {kohavi2017online}
\bibfield{author}{\bibinfo{person}{Ron Kohavi} {and} \bibinfo{person}{Roger
  Longbotham}.} \bibinfo{year}{2017}\natexlab{}.
\newblock \showarticletitle{Online Controlled Experiments and A/B Testing.}
\newblock \bibinfo{journal}{\emph{Encyclopedia of Machine Learning and Data
  Mining}} \bibinfo{volume}{7}, \bibinfo{number}{8} (\bibinfo{year}{2017}),
  \bibinfo{pages}{922--929}.
\newblock


\bibitem[\protect\citeauthoryear{Ma, Zhao, Yi, Yang, Chen, Tang, Hong, and
  Chi}{Ma et~al\mbox{.}}{2020}]%
        {ma2020off}
\bibfield{author}{\bibinfo{person}{Jiaqi Ma}, \bibinfo{person}{Zhe Zhao},
  \bibinfo{person}{Xinyang Yi}, \bibinfo{person}{Ji Yang},
  \bibinfo{person}{Minmin Chen}, \bibinfo{person}{Jiaxi Tang},
  \bibinfo{person}{Lichan Hong}, {and} \bibinfo{person}{Ed~H Chi}.}
  \bibinfo{year}{2020}\natexlab{}.
\newblock \showarticletitle{Off-policy Learning in Two-stage Recommender
  Systems}. In \bibinfo{booktitle}{\emph{The Web Conference 2020}}.
  \bibinfo{publisher}{ACM}, \bibinfo{pages}{463--473}.
\newblock


\bibitem[\protect\citeauthoryear{Oosterhuis and de~Rijke}{Oosterhuis and
  de~Rijke}{2020}]%
        {oosterhuis2020topkrankings}
\bibfield{author}{\bibinfo{person}{Harrie Oosterhuis} {and}
  \bibinfo{person}{Maarten de Rijke}.} \bibinfo{year}{2020}\natexlab{}.
\newblock \showarticletitle{Policy-Aware Unbiased Learning to Rank for Top-k
  Rankings}. In \bibinfo{booktitle}{\emph{SIGIR}}. \bibinfo{publisher}{ACM}.
\newblock


\bibitem[\protect\citeauthoryear{Ovaisi, Ahsan, Zhang, Vasilaky, and
  Zheleva}{Ovaisi et~al\mbox{.}}{2020}]%
        {ovaisi2020correcting}
\bibfield{author}{\bibinfo{person}{Zohreh Ovaisi}, \bibinfo{person}{Ragib
  Ahsan}, \bibinfo{person}{Yifan Zhang}, \bibinfo{person}{Kathryn Vasilaky},
  {and} \bibinfo{person}{Elena Zheleva}.} \bibinfo{year}{2020}\natexlab{}.
\newblock \showarticletitle{Correcting for Selection Bias in Learning-to-rank
  Systems}. In \bibinfo{booktitle}{\emph{WWW}}. \bibinfo{pages}{1863--1873}.
\newblock


\bibitem[\protect\citeauthoryear{Qin and Liu}{Qin and Liu}{2013}]%
        {qin2013introducing}
\bibfield{author}{\bibinfo{person}{Tao Qin} {and} \bibinfo{person}{Tie-Yan
  Liu}.} \bibinfo{year}{2013}\natexlab{}.
\newblock \showarticletitle{Introducing LETOR 4.0 datasets}.
\newblock \bibinfo{journal}{\emph{arXiv preprint arXiv:1306.2597}}
  (\bibinfo{year}{2013}).
\newblock


\bibitem[\protect\citeauthoryear{Radlinski and Craswell}{Radlinski and
  Craswell}{2013}]%
        {radlinski2013optimized}
\bibfield{author}{\bibinfo{person}{Filip Radlinski} {and} \bibinfo{person}{Nick
  Craswell}.} \bibinfo{year}{2013}\natexlab{}.
\newblock \showarticletitle{Optimized Interleaving for Online Retrieval
  Evaluation}. In \bibinfo{booktitle}{\emph{WSDM}}. ACM,
  \bibinfo{pages}{245--254}.
\newblock


\bibitem[\protect\citeauthoryear{Radlinski, Kurup, and Joachims}{Radlinski
  et~al\mbox{.}}{2008}]%
        {radlinski2008does}
\bibfield{author}{\bibinfo{person}{Filip Radlinski}, \bibinfo{person}{Madhu
  Kurup}, {and} \bibinfo{person}{Thorsten Joachims}.}
  \bibinfo{year}{2008}\natexlab{}.
\newblock \showarticletitle{How Does Clickthrough Data Reflect Retrieval
  Quality?}. In \bibinfo{booktitle}{\emph{CIKM}}. ACM, \bibinfo{pages}{43--52}.
\newblock


\bibitem[\protect\citeauthoryear{Schuth, Hofmann, and Radlinski}{Schuth
  et~al\mbox{.}}{2015}]%
        {schuth2015predicting}
\bibfield{author}{\bibinfo{person}{Anne Schuth}, \bibinfo{person}{Katja
  Hofmann}, {and} \bibinfo{person}{Filip Radlinski}.}
  \bibinfo{year}{2015}\natexlab{}.
\newblock \showarticletitle{Predicting Search Satisfaction Metrics with
  Interleaved Comparisons}. In \bibinfo{booktitle}{\emph{SIGIR}}.
  \bibinfo{pages}{463--472}.
\newblock


\bibitem[\protect\citeauthoryear{Wang, Bendersky, Metzler, and Najork}{Wang
  et~al\mbox{.}}{2016}]%
        {wang2016learning}
\bibfield{author}{\bibinfo{person}{Xuanhui Wang}, \bibinfo{person}{Michael
  Bendersky}, \bibinfo{person}{Donald Metzler}, {and} \bibinfo{person}{Marc
  Najork}.} \bibinfo{year}{2016}\natexlab{}.
\newblock \showarticletitle{Learning to Rank with Selection Bias in Personal
  Search}. In \bibinfo{booktitle}{\emph{SIGIR}}. ACM,
  \bibinfo{pages}{115--124}.
\newblock


\bibitem[\protect\citeauthoryear{Wang, Golbandi, Bendersky, Metzler, and
  Najork}{Wang et~al\mbox{.}}{2018a}]%
        {wang2018position}
\bibfield{author}{\bibinfo{person}{Xuanhui Wang}, \bibinfo{person}{Nadav
  Golbandi}, \bibinfo{person}{Michael Bendersky}, \bibinfo{person}{Donald
  Metzler}, {and} \bibinfo{person}{Marc Najork}.}
  \bibinfo{year}{2018}\natexlab{a}.
\newblock \showarticletitle{Position Bias Estimation for Unbiased Learning to
  Rank in Personal Search}. In \bibinfo{booktitle}{\emph{WSDM}}. ACM,
  \bibinfo{pages}{610--618}.
\newblock


\bibitem[\protect\citeauthoryear{Wang, Li, Golbandi, Bendersky, and
  Najork}{Wang et~al\mbox{.}}{2018b}]%
        {wang2018lambdaloss}
\bibfield{author}{\bibinfo{person}{Xuanhui Wang}, \bibinfo{person}{Cheng Li},
  \bibinfo{person}{Nadav Golbandi}, \bibinfo{person}{Michael Bendersky}, {and}
  \bibinfo{person}{Marc Najork}.} \bibinfo{year}{2018}\natexlab{b}.
\newblock \showarticletitle{The LambdaLoss Framework for Ranking Metric
  Optimization}. In \bibinfo{booktitle}{\emph{CIKM}}. ACM,
  \bibinfo{pages}{1313--1322}.
\newblock


\end{thebibliography}
